\documentclass[aps,prx,twocolumn,superscriptaddress,longbibliography,nofootinbib]{revtex4-2}
\pdfoutput=1
\usepackage{amsmath,amssymb,amsthm}
\usepackage{graphicx}
\usepackage{braket}
\usepackage{bm}
\usepackage{booktabs}
\usepackage{xcolor}
\usepackage{microtype}
\usepackage{quantikz}
\usepackage[normalem]{ulem}
\usepackage{comment}
\usepackage[colorlinks=true,allcolors=blue]{hyperref}

\begin{document}

\title{Success Probability of Block-Encoding Methods}

\author{Oumaya Ladhari}
\email{oumaya.laadhari@totalenergies.com}
\affiliation{TotalEnergies, 2 place Jean Millier, 92078 Paris La Défense Cedex, France}
\affiliation{Université de Bordeaux, CNRS, LOMA, UMR 5798, F-33400 Talence, France}
\author{Matthieu Saubanère}
\affiliation{Université de Bordeaux, CNRS, LOMA, UMR 5798, F-33400 Talence, France}
\author{J\'er\'emie Messud}
\affiliation{TotalEnergies, 2 place Jean Millier, 92078 Paris La Défense Cedex, France}

\begin{abstract}
Block-encoding methods apply nonunitary functions $f(\widetilde H)$ of a normalized Hermitian operator on a quantum computer, but only upon a successful postselection, which introduces a shot overhead rarely included in resource estimates. We analyze the success probability $p^{(f)}_{\mathrm{success}}=\sum_n w_n|f(\tilde E_n)|^2$ of three constructions, namely linear combination of unitaries (LCU), qubitization and quantum singular value transformation (QSVT), where $w_n$ are the weights of the input state on the eigenstates of $\widetilde H$. Using the one-dimensional Hubbard model with up to 14 sites (35 qubits), we show that for LCU, the success probability is bounded by the squared ratio between the spectral norm and the $1$-norm of the Pauli decomposition, and that for a Hartree--Fock input it plateaus with system size while the ground-state overlap keeps decreasing. For qubitization, it oscillates with the Chebyshev degree, and favorable degrees can be identified from a few dominant spectral components or predicted from shallow-degree data. For QSVT, the polynomial normalization constant directly sets the success probability, and a tight certified bound balances it against phase-finding robustness. An application to quantum Krylov methods shows a nonuniform shot overhead across Krylov moments. The success probability should therefore be part of the resource estimation of block-encoding-based algorithms.
\end{abstract}

\maketitle

\section{Introduction}
\label{sec:intro}

Block-encoding methods provide a general route for applying nonunitary transformations of an operator $H$ on quantum computers. Such transformations are needed in many applications, such as solving linear systems of equations~\cite{Harrow_2009,Childs_2017,qsvt}, solving partial differential equations~\cite{Childs_2021}, or solving the Schrödinger equation for strongly correlated many-electron systems, e.g., using quantum Krylov subspace methods~\cite{Motta2024,Epperly_2022,oqkd,parrish_mcmahon_2019,stair2019multireferencequantumkrylovalgorithm,Cortes_QuantumKrylov,klymko_2022,Krylovdiagonalization}.
In the latter case, the task is to determine the low-energy spectrum of the Hamiltonian $H$ of the electronic system, in particular its ground-state energy, which can also be targeted by block-encoding-based ground-state preparation and energy estimation~\cite{lin_tong_2020,ge_2019,dong_lin_tong_2022}. Quantum Krylov subspace methods address this problem by constructing a subspace from repeated applications of $H$ to an input state. Because the Hamiltonian $H$ is not unitary, these applications cannot be realized directly on a quantum computer.

Let us first consider a block-encoding of a general Hermitian operator $H$. One constructs a unitary $U$ acting on the system together with an ancilla register $a$, such that a designated block of $U$ (the top-left block, selected by projecting the ancilla onto $\ket{0}_a$) satisfies~\cite{low_hamiltonian_2019}
\begin{equation}
    (\bra{0}_a \otimes I)\, U\, ( \ket{0}_a \otimes I) = \widetilde{H},
    \label{eq:block_encoding_def}
\end{equation}
where
\begin{equation}
    \widetilde{H}=H/\lambda
\end{equation}
is a renormalized version of $H$ such that its spectral norm is less than or equal to one. $\widetilde{H}$ can also include, when necessary, a shift.
The application of $\widetilde{H}$ to an input state $\ket{\psi}$ is recovered when the ancilla register is measured in the $\ket{0}_a$ state. The probability of obtaining this outcome, which we refer to as the success probability, is a central feature investigated in this work.

The Linear Combination of Unitaries (LCU) framework~\cite{childs2012, LCU, low_hamiltonian_2019} provides a constructive way to build such a block-encoding. Writing the Hermitian operator $H$ as a weighted sum of Pauli strings $P_j$,
\begin{equation}
H = \sum_j \alpha_j P_j, 
    \label{eq:Pauli}
\end{equation}
the LCU block-encoding is assembled from two operations acting on an ancilla register $a$ of $n_a$ qubits. A ``prepare'' operator creates a coefficient-weighted superposition on the ancilla,
\begin{equation}
    \mathrm{PREP}\ket{0}_a = \sum_j \sqrt{\tfrac{|\alpha_j|}{\lambda}}\, \ket{j}_a \equiv \ket{\mathrm{PREP}},
    \label{eq:prepare_op}
\end{equation}
and a ``select'' operator applies each Pauli string unitary conditioned on the ancilla index,
\begin{equation}
    \mathrm{SELECT} = \sum_j \ket{j}\!\bra{j}_a \otimes \mathrm{sgn}(\alpha_j)\,P_j ,
    \label{eq:select_op}
\end{equation}
where we write $a=0$ for the measurement outcome in which all ancilla qubits are found in $\ket{0}$.
Together, these operations directly give a block-encoding satisfying Eq.~\eqref{eq:block_encoding_def} through 
\begin{equation}
    U = \mathrm{PREP}^\dagger\, \mathrm{SELECT}\, \mathrm{PREP}.
    \label{eq:lcu_block_encoding}
\end{equation}
An important point is that the necessary unitarity of the preparation operator in Eq.~\eqref{eq:prepare_op} imposes that the renormalization factor
equals the so-called ``LCU 1-norm''
\begin{equation}
    \lambda = \sum_j |\alpha_j|\quad(\text{LCU 1-norm}).
        \label{eq:LCU_1-norm}
\end{equation}

Further refinements of the block-encoding construction give access to increasingly general functions of the Hermitian operator $H$.
For instance, qubitization builds on this same prepare and select pair: combining SELECT with a reflection about the $\ket{\mathrm{PREP}}$ state, $R_\mathrm{PREP} = 2\ket{\mathrm{PREP}}\!\bra{\mathrm{PREP}} - I$, gives the quantum walk operator $W = R_\mathrm{PREP}\,\mathrm{SELECT}$ used here~\cite{low_hamiltonian_2019}.
Powers of $W$ generate Chebyshev polynomials of $\widetilde{H}$,
\begin{equation}
    (\bra{\mathrm{PREP}}\otimes I)\, W^k\, (\ket{\mathrm{PREP}} \otimes I) = T_k(\widetilde{H}),
    \label{eq:chebyshev}
\end{equation}
so that a single block-encoding form gives access to an entire family of transformations indexed by a degree $k$. The LCU case is recovered at $k=1$, since $T_1(x)=x$. 
Finally, Quantum Singular Value Transformation (QSVT)~\cite{low2017,qsvt,martyn2021} generalizes this construction to a broad class of polynomial transformations $P(\widetilde{H})$ of definite parity, renormalized by a factor $C\geq\sup_{x\in[-1,1]}|P(x)|$ so that $|P(x)/C|\le1$ on $[-1,1]$. For a polynomial of degree $d$ and a phase sequence $\Phi=(\phi_0,\phi_1,\dots,\phi_d)$, we define the phase-modulated walk-operator product
\begin{equation}
    \begin{gathered}
    U_\Phi = Z(\phi_0)\,W\,Z(\phi_1)\,W\cdots W\,Z(\phi_d),\\
    Z(\phi)=\mathrm{e}^{\mathrm{i}\phi R_{\mathrm{PREP}}},
    \end{gathered}
    \label{eq:qsvt_equation}
\end{equation}
where the walk operator $W$ is applied $d$ times and $R_{\mathrm{PREP}}$ is the reflection introduced above. Considering real polynomials, so that $P(\widetilde H)$ is Hermitian, QSVT guarantees that there exists a phase sequence $\Phi$ such that~\cite{qsvt}
\begin{equation}
    (\bra{\mathrm{PREP}}\otimes I)\,U_\Phi\,(\ket{\mathrm{PREP}}\otimes I)=\frac{P(\widetilde H)}{C}.
    \label{eq:qsvt_block}
\end{equation}
The phases $\Phi$ are determined classically beforehand from the coefficients of $P/C$~\cite{haah2019,pyqsp,symqsp}. Qubitization and QSVT block-encodings thus give access to different encoded functions $f(\widetilde H)$.

The three constructions discussed above, i.e., LCU, qubitization, and QSVT block-encodings, form the scope of this work.
They correspond to $f=I$ (the identity, $f(x)=x$) for LCU, $f(x)=T_k(x)$ for qubitization, and $f(x)=P(x)/C$ for QSVT.
An important shared property of these three constructions is that the desired application of $f(\tilde H)$ to an input state $\ket{\psi}$ is obtained only upon a successful postselection of the ancilla register. 
The corresponding success probabilities are denoted by
\begin{equation}
    \begin{aligned}
    &p^{(I)}_{\mathrm{success}} &&\text{(LCU)},\\
    &p^{(T_k)}_{\mathrm{success}} &&\text{(qubitization)},\\
    &p^{(P/C)}_{\mathrm{success}} &&\text{(QSVT)},
    \end{aligned}
    \label{eq:psuccess_notation}
\end{equation}
and $p^{(f)}_{\mathrm{success}}$ refers to a generic encoded function. Throughout, lowercase $p$ denotes a success probability, while $P$ denotes a polynomial and $P_j$ a Pauli string. 
For a direct postselection protocol, a success probability $p^{(f)}_{\mathrm{success}}$ implies on average an overhead of $1/p^{(f)}_{\mathrm{success}}$ in the number of circuit repetitions, or shots. 

Understanding the success probability in detail is not straightforward, since it is determined by more than the operator $H$ itself.
Indeed, it depends on the spectral properties of the renormalized operator, the considered function $f(\tilde H)$,
and also the renormalization used to construct the block-encoding and the input state to which the $f(\tilde H)$ is applied. 
We write the spectral decomposition of the Hamiltonian, of its normalized form and of the input state as
\begin{equation}
    \begin{gathered}
    H \ket{E_n} = E_n \ket{E_n}, \qquad \widetilde{H} \ket{E_n} = \tilde{E}_n \ket{E_n},\\
    \ket{\psi} = \sum_n \beta_n \ket{E_n}, \qquad w_n = |\beta_n|^2,
    \end{gathered}
    \label{eq:spectral_decomposition}
\end{equation}
where $E_n$ are the eigenvalues of $H$, ordered so that $E_0$ is the ground-state energy in the half-filled sector to which all input states considered here belong. We denote by $E_{\min}$ and $E_{\max}$ the extremal eigenvalues of $H$ over the full Fock space of the $2N$ system qubits, on which the block-encoding acts. $\tilde{E}_n\in[-1,1]$ corresponds to normalized eigenvalues of $\widetilde{H}$ ($\tilde{E}_n = E_n/\lambda$, or $\tilde{E}_n=(E_n-E_{\max})/\lambda_s$ if a shift is introduced, with $\lambda_s$ the $1$-norm of the shifted decomposition).
Interestingly, the block-encoding implementing an encoded function $f(\widetilde{H})$ succeeds with probability
\begin{equation}
    p^{(f)}_{\mathrm{success}} = \|f(\widetilde{H})\ket{\psi}\|^2 = \sum_n w_n\, |f(\tilde{E}_n)|^2 .
    \label{eq:psuccess_general}
\end{equation}
Equation~\eqref{eq:psuccess_general} provides the common spectral framework used throughout this work: for a given encoded function $f$, the success probability is determined jointly by the spectral weights of the input state, the spectrum of the normalized Hamiltonian, and the normalization required to implement $f$.

Resource estimates for block-encodings of electronic Hamiltonians have focused on reducing the coefficient $1$-norm $\lambda$, through factorized decompositions~\cite{babbush_encoding_2018,von_burg_2021,lee_2021,reiher_2017,berry_2019}, orbital transformations~\cite{koridon_2021} or Hamiltonian partitionings~\cite{loaiza_2023}, in order to reduce the circuit depth; to our knowledge, the shot overhead due to the success probability has usually not been reported. In this article, we analyze this success probability for the three constructions above, using the one-dimensional Hubbard model as a controlled testbed, with the Hartree--Fock (HF) state as main input state, compared with the exact ground state and random states. To reach system sizes beyond those accessible with a dense Hamiltonian, we evaluate the action of $H$ on state vectors directly from its Pauli decomposition, which reduces the memory scaling from $O(4^{n_\mathrm{sys}})$ to $O(2^{n_\mathrm{sys}})$ and allows us to study up to $14$ sites ($28$ system qubits, $35$ qubits including the LCU ancilla register). For each construction we address one question: for LCU, how much the $1$-norm limits the success probability; for qubitization, whose success probability oscillates with the degree $k$, whether favorable degrees can be identified without computing the full spectrum, with an application to quantum Krylov methods; and for QSVT, how the polynomial scaling factor $C$ trades success probability against the robustness of phase finding. For the latter, we use a certified upper bound $C_\mathrm{Markov}$ on the supremum of the target polynomial, obtained from grid samples through the Markov inequality; to our knowledge, such a certificate has not previously been used to normalize QSVT targets.

The article is structured as follows. Section~\ref{sec:main_learnings} summarizes how the LCU $1$-norm, the features of $H$ and the input state determine the success probability. Section~\ref{sec:results} presents the numerical results on the Hubbard model: the computational setup (Sec.~\ref{ssec:hubbard_ham}), LCU (Sec.~\ref{sec:lcu}), qubitization, including degree selection and the application to quantum Krylov (Sec.~\ref{ssec:qubitization}), and QSVT (Sec.~\ref{ssec:qsvt}). We conclude in the last section.

\section{Summary of main findings}
\label{sec:main_learnings}

This section collects the analytical results that organize our findings; the numerical evidence is given in Sec.~\ref{sec:results}. The three constructions differ in encoded function, implementation primitive and normalization (Table~\ref{tab:lcu_qubitization_qsvt}). Within Eq.~\eqref{eq:psuccess_general}, the success probability then depends on the normalization (Sec.~\ref{ssec:dep_norm}), on the features of $H$ and the system size (Sec.~\ref{ssec:dep_system}), and on the input state (Sec.~\ref{ssec:dep_state}).
\begin{table*}[tbp]
\centering
\caption{Comparison of the block-encoding constructions considered in this work.}
\label{tab:lcu_qubitization_qsvt}
\renewcommand{\arraystretch}{1.25}
\begin{tabular}{llll}
\toprule
\textbf{Property} &
\textbf{LCU} &
\textbf{Qubitization} &
\textbf{QSVT} \\
\midrule
Encoded function $f(x)$ &
$x$ &
$T_k(x)$ &
$P(x)/C$ ($P$ real, definite parity) \\
Primitive &
PREP-SELECT &
Walk operator $W$ &
QSVT phase sequence \\
Main control parameter &
LCU 1-norm $\lambda$ &
Degree $k$ &
Polynomial $P$ and scaling $C$ \\
Success probability &
$p^{(I)}_{\mathrm{success}}=\sum_n w_n \tilde{E}_n^2$ &
$p^{(T_k)}_{\mathrm{success}}=\sum_n w_n T_k(\tilde{E}_n)^2$ &
$p^{(P/C)}_{\mathrm{success}} = \sum_n w_n |P(\tilde{E}_n)|^2/C^2$ \\
\bottomrule
\end{tabular}
\end{table*}

\subsection{Dependence of the success probability on the LCU 1-norm}
\label{ssec:dep_norm}
We consider a Pauli decomposition as in Eq.~\eqref{eq:Pauli}, and use the notations above, together with the spectral norm of $H$ (shifted if necessary),
\begin{equation}
    \Omega = \| H\|, \qquad \lambda \ge \Omega.
    \label{eq:spectral-norm}
\end{equation}

For a fixed input state $\ket{\psi}$ as in Eq.~\eqref{eq:spectral_decomposition}, the success probability of a general block-encoding [Eq.~\eqref{eq:block_encoding_def}] increases as the renormalization factor $\lambda$ (which in this general setting is not necessarily the LCU $1$-norm) approaches the spectral norm $\Omega$ of $H$ [Eq.~\eqref{eq:spectral-norm}]. Replacing the LCU $1$-norm [Eq.~\eqref{eq:LCU_1-norm}] by the tighter $\Omega$ therefore provides an upper benchmark for the achievable success probability~\cite{babbush_encoding_2018}. The corresponding success probabilities under the two renormalization choices are
\begin{equation}
    \begin{aligned}
    p^{(I)}_{\mathrm{success},\,1\text{-norm}} &= \frac{1}{\lambda^2} \sum_n w_n E_n^2,\\
    p^{(I)}_{\mathrm{success},\,\text{spectral}} &= \frac{1}{\Omega^2} \sum_n w_n E_n^2,
    \end{aligned}
\end{equation}
where $E_n$ are the eigenvalues of $H$ before renormalization. This leads to:
\begin{equation}
    \frac{p^{(I)}_{\mathrm{success},\,\text{spectral}}}{p^{(I)}_{\mathrm{success},\,1\text{-norm}}} = \frac{\lambda^2}{\Omega^2} \geq 1 .
\end{equation}
A standard technique is to pin the spectral maximum to zero~\cite{low_hamiltonian_2019}:
\begin{equation}
\label{eq:spectral_shift}
    H_s\equiv H-E_{\mathrm{max}}I,
    \qquad E_{\mathrm{max}}=\max_n E_n.
\end{equation}
After this shift all eigenvalues lie in $[E_{\min}-E_{\max},0]$ and
$\Omega_s = E_{\max}-E_{\min}$, the spectral norm of $H_s$ over the full Fock space. For the Hubbard model, $E_{\min}=E_0$ at weak coupling, but beyond a crossover ($U=3t$ for $N=2$, between $t$ and $2t$ for $N=4$) the global minimum lies in a sector with fewer electrons, which avoids the double-occupancy cost, so that $E_{\min}<E_0$. Since all input states belong to a single particle-number sector, the normalization could in principle be tightened by symmetry-adapted shifts that leave this sector unchanged, such as the block-invariant symmetry shift of Ref.~\cite{loaiza_bliss_2023}.
The spectral norm $\Omega_s$ is the smallest normalization compatible with $\|\widetilde{H}\|_2\le1$, leading to $\|\widetilde{H}\|_2=1$.
We thus have $p^{(I)}_{\mathrm{success},\,\text{spectral}} \geq p^{(I)}_{\mathrm{success},\,1\text{-norm}}$ and $p^{(I)}_{\mathrm{success},\,\text{spectral}}=1$ iff $\ket{\psi}=\ket{E_{\min}}$ (up to a phase).

The PREP-SELECT construction [Eq.~\eqref{eq:lcu_block_encoding}] related to the LCU block-encoding specific form, however, constrains the usage of the LCU $1$-norm, i.e., $\lambda_s=\sum_j|\alpha_j|$ of the shifted decomposition, because the PREP operator must be unitary. The physically realized success probability is therefore always
\begin{equation}
    \begin{aligned}
    p^{(I)}_{\mathrm{success}}&=\frac{\|H_s\ket{\psi}\|^2}{\lambda_s^2}
    =\left(\frac{\Omega_s}{\lambda_s}\right)^{2}\frac{\|H_s\ket{\psi}\|^2}{\Omega_s^2}\\
    &\le \left(\frac{\Omega_s}{\lambda_s}\right)^{2}\equiv p_{\mathrm{bound}},
    \end{aligned}
        \label{eq:prob_upper_bound}
\end{equation}
where the first factor, $(\Omega_s/\lambda_s)^2\le1$, is the loss due to the LCU $1$-norm, and the second, $\|H_s\ket{\psi}\|^2/\Omega_s^2$, is the success probability that a block-encoding normalized by the spectral norm would reach; it depends only on the input state. Improving the success probability can only be achieved by reducing $\lambda_s$ or by the choice of $\ket{\psi}$.

For qubitization [Eq.~\eqref{eq:chebyshev}], the realized value uses the LCU $1$-norm, $\tilde{E}_n=(E_n-E_{\max})/\lambda_s$,
\begin{equation}
    p^{(T_k)}_{\mathrm{success}} = \sum_n w_n \, T_k\!\left(\frac{E_n - E_{\max}}{\lambda_s}\right)^{\!2}.
    \label{eq:P_qubit_normalisation}
\end{equation}
Unlike the LCU case, the LCU $1$-norm does not affect the qubitization success probability in a clearly predictable way because  $T_k$ is nonlinear. The same holds for the QSVT success probability. For QSVT, with target polynomial $P(x)$ and scaling factor $C$, the implemented polynomial is $P(x)/C$, giving
\begin{equation}
    p^{(P/C)}_{\mathrm{success}} = \left\| \frac{P(\widetilde{H})}{C} \ket{\psi} \right\|^2= \frac{1}{C^2} \sum_n w_n \, |P(\tilde{E}_n)|^2 .
    \label{eq:p_qsvt}
\end{equation}
As with qubitization, $P$ is nonlinear, so the $H$ renormalization does not affect the QSVT success probability in a clearly predictable way. In addition, QSVT introduces its own scaling factor 
\begin{equation}
C\geq C_\mathrm{opt}=\sup_{x\in[-1,1]}|P(x)|,
\end{equation}
required for $|P(x)/C|\leq1$ on $[-1,1]$; any bound looser than the supremum reduces the success probability as $1/C^2$ (Sec.~\ref{ssec:qsvt}).

In most of the remaining analysis, we work with the shifted and normalized Hamiltonian, whose eigenvalues lie in $[-1,0]$. This one-sided spectrum simplifies the analysis of the Chebyshev transformations and provides a convenient setting in which to derive bounds of the success probability.

\subsection{Dependence of the success probability on system features}
\label{ssec:dep_system}
Beyond the choice of normalization, the features of $H$ and the system size affect both the spectral distribution and the normalization itself.

In the Hubbard model, the important feature of $H$ is the interaction strength. When it grows, the coefficient $1$-norm $\lambda_s$ accumulates contributions from every term that the interaction activates, whereas the spectral norm $\Omega_s$ tracks only the largest spectral gap. The bound $(\Omega_s/\lambda_s)^2$ of Eq.~\eqref{eq:prob_upper_bound} therefore decreases with $U/t$, from its free-fermion value ($1$ for $N=2$, $\approx0.63$ for $N=14$, tending to $(4/(\pi+2))^2\approx0.61$ at large $N$) to $4/9$ for every $N$: at large $U$, $\Omega_s=E_{\max}-E_{\min}\to NU$ while $\lambda_s\to 3NU/2$ (Sec.~\ref{sec:lcu}).
A central and initially counterintuitive finding concerns the dependence on system size: for the Hubbard systems and structured input states considered, the success probability shows no systematic decay with system size, both for LCU (Sec.~\ref{sec:lcu}) and QSVT (Sec.~\ref{ssec:qsvt_size}) block-encodings. Although $\lambda_s$ increases with system size, the spectral distribution of the input state changes with it, and these two effects compensate: for the HF input state, $p^{(I)}_{\mathrm{success}}$ settles onto a plateau rather than decaying toward zero. In the Hubbard model this plateau is reached already at $N=4$: the success probability separates most sharply between $N=2$ and $N\geq4$ in the intermediate-correlation crossover, while the curves for $N\geq4$ are nearly indistinguishable and converge to the same large-$U/t$ value (Fig.~\ref{fig:combined}(b)).

\subsection{Dependence of the success probability on the input state and ground-state overlap}
\label{ssec:dep_state}
The input state itself plays a role that differs between the LCU and qubitization block-encodings. For the LCU block-encoding, the success probability is maximized by the eigenvector $\ket{E_{\min}}$, which after the spectral shift carries the eigenvalue of largest magnitude: $p^{(I)}_{\mathrm{success}}(\ket{\psi}) \leq (\Omega_s/\lambda_s)^2 = p^{(I)}_{\mathrm{success}}(\ket{E_{\min}})$ for every normalized $\ket{\psi}$. Within the half-filled sector the maximizer is the ground state $\ket{E_0}$, which reaches the bound when $E_0=E_{\min}$ and lies slightly below it otherwise (Sec.~\ref{sec:lcu}).
For qubitization, the state dependence is qualitatively richer. Writing $\tilde E_n=\cos\theta_n$, so that $T_k(\tilde E_n)=\cos(k\theta_n)$, and using $\cos^2(k\theta_n)=[1+\cos(2k\theta_n)]/2$, the qubitization success probability becomes
\begin{equation}
    p^{(T_k)}_{\mathrm{success}}=\frac12+\frac12\sum_n w_n\cos(2k\theta_n).
\end{equation}
Its average over degrees is therefore
\begin{equation}
    \overline{p^{(T_k)}_{\mathrm{success}}}\equiv\lim_{K\to\infty}\frac1K\sum_{k=1}^{K}p^{(T_k)}_{\mathrm{success}}
    =\frac12+\frac12\sum_{n:\,\theta_n=\pi}w_n ,
\end{equation}
since $(1/K)\sum_{k}\cos(2k\theta_n)\to0$ unless $\theta_n$ is a multiple of $\pi$~\cite{weyl_gleichverteilung_1916,kuipers_uniform_1974}. The input state controls the fluctuations around this average. Writing $p^{(T_k)}_{\mathrm{success}}=[1+\sum_n w_n\cos(2k\theta_n)]/2$, the variance over $k$ is $(1/8)\sum_n w_n^2$ for distinct $\theta_n$, so the inverse participation ratio $\sum_n w_n^2$ directly sets the amplitude of the oscillations. A state with broad spectral support, e.g., a random state, for which $\sum_n w_n^2\approx2/(4^N+1)$, remains close to $1/2$, whereas a state concentrated on a few eigenstates (large $\sum_n w_n^2$) explores a much broader range of values even at large $k$. This oscillation makes the degree an additional lever: for concentrated input states, favorable degrees can be identified from a few dominant spectral components, or predicted from shallow-degree data without classical diagonalization (Sec.~\ref{sssec:a_priori}). In the Hubbard model, the HF input state exhibits such a concentrated spectral structure across the system sizes studied, whereas a random state has broad spectral support.

The success probability should also be distinguished from the fidelity of the input state, i.e., its overlap $|\braket{\psi_{\mathrm{HF}}|E_0}|^2$ with the ground state. The fidelity measures alignment with a single eigenvector, whereas $p^{(I)}_{\mathrm{success}}$ is a weighted average over the whole spectrum. As a result, $p^{(I)}_{\mathrm{success}}$ is quasi-independent of the fidelity of the HF input: it plateaus with system size while the overlap keeps falling (Sec.~\ref{sec:lcu}). This differs from filtering-based ground-state preparation, where the success probability is set by the ground-state overlap~\cite{lin_tong_2020}. 

\subsection{Success probability in resource estimation}
In a direct postselection protocol, the shot overhead per accepted outcome is $1/p^{(f)}_{\mathrm{success}}$, and when the block-encoding is applied repeatedly, e.g., to the states spanning a Krylov subspace, this overhead accumulates over the course of the algorithm~\cite{Kirby_2023,Lee_2024,Zhang_2024}. Amplitude amplification and its oblivious and fixed-point variants reduce the number of repetitions to $O(1/\sqrt{p^{(f)}_{\mathrm{success}}})$~\cite{brassard_2002,berry_2015,yoder_2014}, at the cost of deeper circuits. A method with shallower circuits but a lower success probability may therefore be more costly overall than a deeper one with a higher acceptance probability, so the success probability should be reported alongside circuit depth, gate count and qubit count, which are the quantities usually estimated~\cite{reiher_2017,kivlichan_2020,campbell_2022}.

\section{Numerical results}
\label{sec:results}

We now turn to the numerical characterization of the success probability for the three methods considered here: LCU, qubitization and QSVT. As a controlled testbed, we use the one-dimensional Hubbard model. This model is well suited for this analysis: it is amenable to exact diagonalization for validation, its correlation strength is tunable through a single ratio, and it admits known analytic limits against which the success probability can be checked.

\begin{table}[t]
\centering
\caption{Summary of the numerical setup used throughout the study.}
\label{tab:numerical_setup}
\renewcommand{\arraystretch}{1.15}
\begin{tabular}{@{}lp{0.58\columnwidth}@{}}
\toprule
\textbf{Quantity} & \textbf{Setting} \\
\midrule
Model & 1D Hubbard chain at half filling \\
Hopping & $t=1$ \\
Interaction & varied through $U/t$ \\
System size & $N=2$--$14$ sites \\
System qubits & $n_{\mathrm{sys}}=2N$ \\
LCU terms & $7N-3$ \\
Spectral shift & $H_s=H-E_{\max}I$ \\
LCU normalization & coefficient $1$-norm $\lambda_s$ \\
Input states & HF, ground state, random \\
Qubitization & $T_k(\widetilde H)$ \\
QSVT & definite-parity Chebyshev targets \\
Largest system & 28 system qubits + 7 ancilla = 35 qubits \\
Validation & Dense matrix for $N\leq6$; matrix-free beyond \\
\bottomrule
\end{tabular}
\end{table}

\subsection{The Hubbard model and computational setup}
\label{ssec:hubbard_ham}

We study the one-dimensional Hubbard model on a chain of $N$ sites at half filling~\cite{hubbard_1963,Dagotto_1994,lieb_wu_1968,essler_2005},
\begin{equation}
\label{eq:hubbard}
    H = -t\sum_{\langle i,j\rangle,\sigma}
        \!\bigl(\hat{c}^\dagger_{i,\sigma}\hat{c}_{j,\sigma}+\mathrm{h.c.}\bigr)
    + U\sum_i\hat{n}_{i,\uparrow}\hat{n}_{i,\downarrow},
\end{equation}
where \(t\) is the nearest-neighbor hopping amplitude, \(U\) is the on-site interaction strength, \(c_{i,\sigma}^{\dagger}\) and \(c_{i,\sigma}\) are fermionic creation and annihilation operators, and
\begin{equation}
n_{i,\sigma}=c_{i,\sigma}^{\dagger}c_{i,\sigma}
\end{equation}
is the occupation operator. The ratio $U/t$ controls the crossover from the weakly interacting regime ($t\gg U$) to the strongly correlated Mott insulating regime with localized electrons ($U\gg t$). Throughout this work, we set \(t=1\) and vary \(U/t\).

We map the fermionic operators to $n_\mathrm{sys}=2N$ qubits using the Jordan--Wigner transformation~\cite{jordan_wigner_1928}, with the spin-up orbitals ordered first. For the nearest-neighbor hopping terms, the Jordan--Wigner strings cancel between adjacent sites, yielding
\begin{equation}
\label{eq:hubbard_pauli}
    \begin{split}
    H ={}&
    -\frac{t}{2}\sum_\sigma\sum_{i=0}^{N-2}
     \bigl(X_{i+\delta_\sigma}X_{i+1+\delta_\sigma}+Y_{i+\delta_\sigma}Y_{i+1+\delta_\sigma}\bigr)\\
    &+\frac{NU}{4}I
    -\frac{U}{4}\sum_{i=0}^{N-1}(Z_i+Z_{N+i})
    +\frac{U}{4}\sum_{i=0}^{N-1}Z_iZ_{N+i}.
    \end{split}
\end{equation}
This decomposition contains $7N-3$ Pauli strings, from which the LCU $1$-norm is obtained directly. We use the spectral shift of Eq.~\eqref{eq:spectral_shift}, so that the normalized eigenvalues lie in $[-1,0]$.

The PREP operator loads the amplitudes $\sqrt{|\alpha_j|/\lambda}$ onto an ancilla register of width $n_a=\lceil\log_2(7N-3)\rceil$ using a Grover--Rudolph binary tree of rotation angles constructed from partial sums of the coefficient distribution~\cite{grover_2002}. The sign factors $\mathrm{sgn}(\alpha_j)$ are incorporated into SELECT rather than PREP.
SELECT applies each signed Pauli string $\mathrm{sgn}(\alpha_j)P_j$ controlled on the ancilla index $\ket{j}_a$. Each term is implemented as a multicontrolled Pauli controlled by the ancilla index, with the control complexity determined by the ancilla width $n_a$. The compiled PREP-SELECT-PREP$^\dagger$ circuit is validated at two levels.
First, for $N \leq 6$, where the dense Hamiltonian remains computationally accessible, the projected circuit block agrees with the explicitly constructed $\widetilde{H}$ to machine precision ($\sim10^{-15}$). Second, beyond this range, the exact HF success probability, evaluated matrix-free by applying $H_s$ term by term to $\ket{\psi_{\mathrm{HF}}}$, $p^{(I)}_{\mathrm{success}}(\ket{\psi_{\mathrm{HF}}})=\|H_s\ket{\psi_\mathrm{HF}}\|^2/\lambda_s^2$ is compared with the shot-based estimate $\hat p^{(I)}_{\mathrm{success}}(\ket{\psi_{\mathrm{HF}}})$, i.e., the fraction of shots in which the ancilla register is measured in $\ket{0}_a$ when the compiled circuit is applied to $\ket{\psi_{\mathrm{HF}}}$; the two agree within the binomial standard error $\sqrt{p^{(I)}_{\mathrm{success}}\big(1-p^{(I)}_{\mathrm{success}}\big)/N_\mathrm{shots}}$, with $N_\mathrm{shots}=20000$, which validates the matrix-free implementation used for $N > 6$.
The numerical setup is summarized in Table~\ref{tab:numerical_setup}.

\subsection{LCU success probability}
\label{sec:lcu}

\subsubsection{Ground state as input state}

\begin{figure*}[tbp]
    \centering
    \begin{minipage}[t]{0.49\textwidth}
        \centering
        \textbf{(a)}\\[2pt]
        \includegraphics[width=\linewidth]{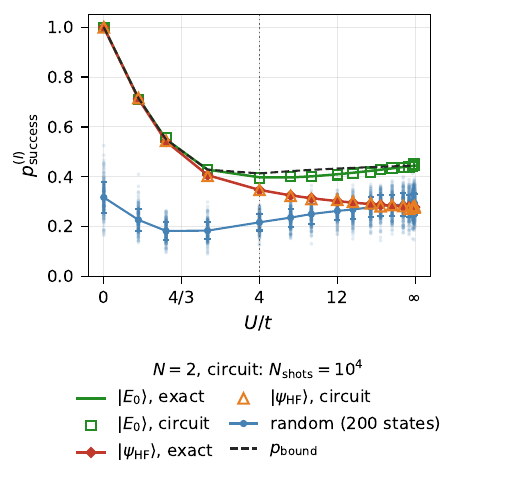}
    \end{minipage}\hfill
    \begin{minipage}[t]{0.49\textwidth}
        \centering
        \textbf{(b)}\\[2pt]
        \includegraphics[width=\linewidth]{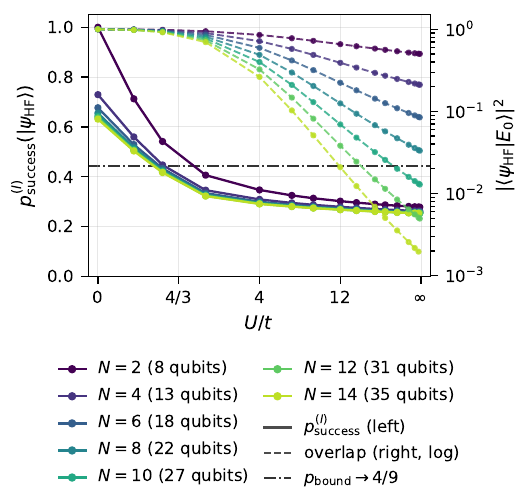}
    \end{minipage}
    \caption{LCU success probability $p^{(I)}_{\mathrm{success}}$ versus $U/t$ (axis scaled as $U/(U+4t)$). \textbf{(a)} $N=2$, for the half-filled ground state $\ket{E_0}$ (green) and the HF state (red): exact value $(\Omega_s/\lambda_s)^2\|H_s\ket{\psi}\|^2/\Omega_s^2$ (lines) and compiled PREP-SELECT circuit sampled with $N_\mathrm{shots}=10^4$ (open markers). The dashed line is $p_\mathrm{bound}=(\Omega_s/\lambda_s)^2$, reached by $\ket{E_0}$ only for $U\lesssim3t$, where the global minimum of $H$ lies at half filling. Blue: random input states (mean $\pm$ standard deviation over 200 states, individual states shown as dots). In the circuit simulations the input state is injected directly as the initial statevector of the simulator. \textbf{(b)} HF input state for $N=2$--$14$ (color): $p^{(I)}_{\mathrm{success}}(\ket{\psi_{\mathrm{HF}}})$ (solid, left axis) and HF--ground-state (GS) overlap $|\braket{\psi_{\mathrm{HF}}|E_0}|^2$ (dashed, right axis, log scale). The dash-dotted line marks the large-$U$ limit $p_{\mathrm{bound}}\to4/9$.}
    \label{fig:combined}
\end{figure*}

The LCU block-encoding success probability obeys the state-independent upper bound in Eq.~\eqref{eq:prob_upper_bound}, which is controlled entirely by the ratio $\Omega_s/\lambda_s$ with $p_{\mathrm{bound}} = (\Omega_s/\lambda_s)^2$, and two limits fix its range. At $U=0$, the spectrum is symmetric $E_{\min}=-E_{\max}$ and the hopping strings contribute $2t(N-1)$ to the $1$-norm, while the constant introduced by the shift contributes $E_{\max}$:
\begin{equation}
\label{eq:pbound_u0}
    \begin{gathered}
    \Omega_s\big|_{U=0}=2E_{\max}, \qquad \lambda_s\big|_{U=0}=2t(N-1)+E_{\max},\\
    p_{\mathrm{bound}}\big|_{U=0}=\left(\frac{2E_{\max}}{2t(N-1)+E_{\max}}\right)^{2} .
    \end{gathered}
\end{equation}
For $N=2$ the two scales coincide ($\Omega_s=\lambda_s=4t$) and $p_{\mathrm{bound}}=1$, but for larger chains the hopping strings do not all saturate the operator norm simultaneously and the bound falls below unity: $p_{\mathrm{bound}}\approx0.73$, $0.68$, $0.66$ and $0.63$ for $N=4$, $6$, $8$ and $14$. These values converge to a constant at large $N$: at $U=0$, $E_{\max}=2\sum_{\varepsilon_q>0}\varepsilon_q$ with $\varepsilon_q=-2t\cos\!\big(q\pi/(N+1)\big)$, so that $E_{\max}\to4tN/\pi$, $\Omega_s=2E_{\max}\to8tN/\pi$ and $\lambda_s\to2tN+4tN/\pi$, giving $p_{\mathrm{bound}}\big|_{U=0}\to\left(4/(\pi+2)\right)^2\approx0.61$.
In the strong-coupling limit the two scales separate. For $U\gg t$, the highest-energy configuration is the fully doubly occupied state, which gives $E_{\max}=NU$, while $|E_{\min}|=O(t)$ is negligible on the scale of $U$. For the shifted Pauli decomposition of Eq.~\eqref{eq:hubbard_pauli}, the coefficients are:
\begin{equation}
\label{eq:largeU_scales}
    \begin{aligned}
    \Omega_s &\xrightarrow{U \gg t} NU,\\
    \lambda_s &= \underbrace{2t(N-1)}_{\text{hopping}} + \underbrace{\tfrac{NU}{2}}_{Z} + \underbrace{\tfrac{NU}{4}}_{ZZ} + \underbrace{\tfrac{3NU}{4}}_{\text{const}}
    \xrightarrow{U \gg t} \tfrac{3NU}{2},
    \end{aligned}
\end{equation}
and therefore
\begin{equation}
\label{eq:pbound_largeU}
    \frac{\Omega_s}{\lambda_s}\xrightarrow{U\gg t}\frac{2}{3}, \qquad p_{\mathrm{bound}}\to\frac{4}{9}\approx0.44 .
\end{equation}
Here the $Z$ and $ZZ$ terms contribute $NU/2$ and $NU/4$ respectively, while the constant term, $NU/4$ before shifting, contributes $3NU/4$ in magnitude after the spectral shift, since $E_{\max}\to NU$ sends it to $NU/4 - E_{\max} \to -3NU/4$. The limit $4/9$ is therefore independent of both $N$ and $U/t$. The decrease of the bound with $U/t$, from its free-fermion value to $4/9$, is the decomposition overhead of the $1$-norm: the Pauli strings per site partially cancel in the operator norm but count additively in $\lambda_s$. Unlike the spectral norm, the $1$-norm depends on the orbital representation used for the Pauli decomposition, so that $p_\mathrm{bound}$ can be increased a priori by optimizing this representation~\cite{koridon_2021}; the site and molecular-orbital bases are compared below.
After the spectral shift, the state of lowest energy $E_{\min}$ carries the eigenvalue of largest magnitude of $H_s$, $|E_{\min}-E_{\max}|=\Omega_s$, and therefore saturates the bound. The half-filled ground state saturates it only when $E_0=E_{\min}$, i.e., for $U<3t$ ($N=2$) and $U\lesssim t$ ($N=4$). Beyond, $p^{(I)}_{\mathrm{success}}(\ket{E_0})=(\Omega_s/\lambda_s)^2(E_{\max}-E_0)^2/(E_{\max}-E_{\min})^2$ falls slightly below $p_\mathrm{bound}$, by at most $\approx8\%$ for $N\le4$ (minimum $0.92\,p_\mathrm{bound}$ at $U\approx6t$ for $N=4$), and returns to it at large $U$, where $E_0$ and $E_{\min}$ are both $O(t)$ against $E_{\max}=NU$. As a consistency check, Fig.~\ref{fig:combined}(a) compares, for $N=2$, the success probability computed from $(\Omega_s/\lambda_s)^2\,\|H_s\ket{\psi}\|^2/\Omega_s^2$ with the one measured on the compiled PREP-SELECT circuit, for the half-filled ground state and for the HF state; the two agree, confirming that the circuit realizes the $1$-norm normalization.

\subsubsection{HF as input state}
We use the HF input state $\ket{\psi_\mathrm{HF}}$, taken here as the single Slater determinant minimizing the energy for the noninteracting problem~\cite{szabo_modern_qchem}. In the site basis it is a superposition of computational basis states. At $U=0$ the HF approximation is exact ($\ket{\psi_\mathrm{HF}} = \ket{E_0}$), so $p^{(I)}_{\mathrm{success}}(\ket{\psi_\mathrm{HF}})=p_{\mathrm{bound}}$, which equals $1$ only for $N=2$ (Eq.~\eqref{eq:pbound_u0}). For $U > 0$, the exact ground state acquires correlations beyond a single Slater determinant~\cite{Dagotto_1994}. Expanding $\ket{\psi_\mathrm{HF}} = \sum_n \beta_n\ket{E_n}$ in the exact eigenbasis of Eq.~\eqref{eq:spectral_decomposition} (with $|\beta_0|^2 < 1$ for $U > 0$ once correlations make the HF input state different from the exact ground state):
\begin{equation}
\label{eq:p_hf_expand}
    p^{(I)}_{\mathrm{success}}(\ket{\psi_\mathrm{HF}})
    = \sum_n w_n \tilde{E}_n^2
    < \tilde{E}_0^2 = p_{\mathrm{bound}}.
\end{equation}
The difference $p_{\mathrm{bound}} - p^{(I)}_{\mathrm{success}}(\ket{\psi_\mathrm{HF}})$ therefore reflects electron correlation, but it is not identical to the loss of ground-state overlap: it is a spectrally weighted measure of how the HF spectral weight is distributed away from the extremal eigenvalue. 

In order to study the effect of the basis on the success probability, we also computed the same metric after a change of basis from the site basis to the molecular-orbital (MO) basis. In the MO basis, the HF state is represented as a single computational-basis occupation string (e.g., $\ket{\psi_\mathrm{HF}} = \ket{1010}$ for $N=2$), requiring only single-qubit $X$ gates for state preparation. However, this simplification does not improve LCU success probability, because the relevant cost is controlled by the Hamiltonian decomposition rather than by state-preparation complexity alone. In the site basis the interaction is local, with fewer Pauli strings contributing to the $1$-norm. The orbital rotation to the MO basis introduces additional nonlocal terms with no site-basis counterpart, so the MO interaction has more terms and we have $\lambda_{\mathrm{MO}}>\lambda_{\mathrm{site}}$, consistent with the observation that localized orbitals reduce the $1$-norm of molecular Hamiltonians~\cite{koridon_2021}.
Since $\Omega_s$ is basis independent:
\begin{equation}
\label{eq:pbound_basis}
    \begin{aligned}
    p_{\mathrm{bound}}^{\mathrm{site}}
    &= (\Omega_s/\lambda_{s,\mathrm{site}})^2
    > (\Omega_s/\lambda_{s,\mathrm{MO}})^2\\
    &= p_{\mathrm{bound}}^{\mathrm{MO}}
    \quad \forall\, U/t > 0.
    \end{aligned}
\end{equation}
Thus, for the success probability considered here, the optimal basis is not necessarily the one that minimizes state-preparation complexity, but rather one that minimizes the ratio $\lambda_s/\Omega_s$.

\subsubsection{Overlap between the ground state and the HF state}

Unlike the overlap $|\beta_0|^2$, which depends only on the ground-state component of $\ket{\psi_\mathrm{HF}}$, $p^{(I)}_{\mathrm{success}}(\ket{\psi_{\mathrm{HF}}})$ of Eq.~\eqref{eq:p_hf_expand} is a second moment over the whole HF spectral distribution. As $U$ increases, the interaction inflates $\lambda_s$ relative to $\Omega_s$: $p_{\mathrm{bound}} = (\Omega_s/\lambda_s)^2$ decreases, compressing all normalized eigenvalues $\tilde{E}_n$ uniformly. The success probability is therefore affected not only by the loss of ground-state weight but also by the rescaling of the entire spectrum induced by the normalization. The two quantities can therefore differ substantially: in Fig.~\ref{fig:combined}(b) we can see that already at $N=2$, $U=4t$ the HF state retains a substantial overlap $|\beta_0|^2\approx0.85$ with the exact ground state, yet $p^{(I)}_{\mathrm{success}}(\ket{\psi_{\mathrm{HF}}})$ has dropped to $\approx0.35$. Since the shot overhead scales as $1/p^{(I)}_{\mathrm{success}}(\ket{\psi_{\mathrm{HF}}})$, this is a more severe requirement than the overlap alone would suggest. More generally, the success probability is quasi-independent of the fidelity of the input state: at large $U$, $p^{(I)}_{\mathrm{success}}(\ket{\psi_{\mathrm{HF}}})$ plateaus at $\approx0.25$--$0.26$ for $N\geq4$, while the HF--GS overlap keeps falling with $N$ (Fig.~\ref{fig:combined}(b)).
For our purposes, this means that the shot overhead of the LCU block-encoding with a HF input remains bounded, over all system sizes studied ($N\le14$), even in the strongly correlated regime where the HF state becomes a poor approximation of the ground state, while the bound itself stays bounded away from zero in both limits, tending to $(4/(\pi+2))^2\approx0.61$ at $U=0$ for large $N$ and to $4/9$ for every $N$ as $U\to\infty$. The success probability therefore does not penalize the use of a simple HF input state in larger systems; conversely, it cannot serve as a proxy for the fidelity of the input state, which must be assessed separately.

\subsection{Qubitization success probability}
\label{ssec:qubitization}

Whereas LCU block-encoding applies $\widetilde H$ directly, qubitization realizes the Chebyshev transformation $T_k(\widetilde H)$. The corresponding success probability is therefore degree dependent:
\begin{equation}
\label{eq:pk_qubit}
    p^{(T_k)}_{\mathrm{success}} = \|T_k(\widetilde H)\ket{\psi}\|^2 = \sum_n w_n\, T_k(\tilde{E}_n)^2
\end{equation}
Writing each eigenvalue as $\tilde{E}_n=\cos\theta_n$ thus
\begin{equation}
\theta_n=\arccos(\tilde{E}_n),
\end{equation}
the Chebyshev identity $T_k(\cos\theta_n)=\cos(k\theta_n)$ turns Eq.~\eqref{eq:pk_qubit} into a weighted sum of oscillating terms,
\begin{equation}
\label{eq:pk_spectral}
    p^{(T_k)}_{\mathrm{success}} = \sum_n w_n\,\cos^2(k\theta_n).
\end{equation}
Two features of Eq.~\eqref{eq:pk_spectral} are particularly relevant to the analysis that follows. First, as $U$ increases, the LCU $1$-norm grows faster than the spectral norm, compressing the normalized spectrum into $[-\Omega_s/\lambda_s,0]$, which tends to $[-2/3,0]$ at large $U$ (Sec.~\ref{sec:lcu}).
Second, at fixed $U$ the probability oscillates with $k$: a single term of Eq.~\eqref{eq:pk_spectral} vanishes when
\begin{equation}
\label{eq:pk_node}
    k\theta_n = \left(m+\tfrac12\right)\pi, \quad m\in\mathbb{Z},
\end{equation}
so $p^{(T_k)}_{\mathrm{success}}$ can become small when the dominant spectral components lie near such nodes. It is this second feature that might make the choice of degree an operational lever.
\begin{figure*}[tbp]
    \includegraphics[width=0.84\textwidth]{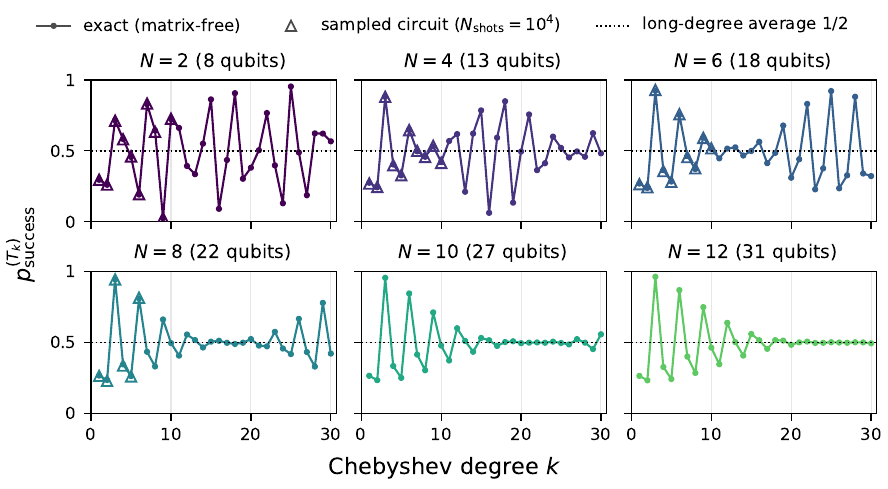}
    \caption{Qubitization success probability $p^{(T_k)}_{\mathrm{success}}$ as a function of the Chebyshev degree $k$ at $U/t=16$ for the HF input state and $N=2$--$12$ sites (color as in Figs.~\ref{fig:combined}(b) and~\ref{fig:combined_qsvt}). Lines and dots show exact values, computed without sampling from the matrix-free action of $H$ (Sec.~\ref{ssec:hubbard_ham}). Open triangles show sampled-circuit estimates with $N_\mathrm{shots}=10^4$, available for $k\le10$ ($N\le6$) and $k\le6$ ($N=8$); their binomial standard deviations ($\le0.005$) are smaller than the markers. The sampled estimates agree with the exact values within two binomial standard deviations (rms deviation $\approx0.005$). The dotted black line marks the long-degree average $1/2$. The oscillatory dependence on $k$ produces degrees with both high and near-vanishing success probabilities.}
    \label{fig:pk_vs_k}
\end{figure*}
At $k=1$ the Chebyshev transformation is the identity, $T_1(\widetilde H)=\widetilde H$, so
\begin{equation}
\label{eq:k1_lcu}
    p^{(T_1)}_{\mathrm{success}} = \|\widetilde H\ket{\psi}\|^2 = p^{(I)}_{\mathrm{success}},
\end{equation}
for every input state and every $U/t$: the LCU success probability is the $k=1$ slice of the qubitization framework. The features specific to qubitization (oscillation with $k$, nodes, and the resulting freedom in choosing the degree) emerge for $k>1$, and make $k$ an additional feature that affects resource estimation: different degrees of the same walk can have substantially different shot overheads.

\subsubsection{Selecting the Chebyshev polynomial degree}
\label{sssec:a_priori}

Because $p^{(T_k)}_{\mathrm{success}}$ oscillates with $k$, a blindly chosen degree can fall near a node, as expressed in Eq.~\eqref{eq:pk_node}, where the success probability can become very small. For the HF input state, whose spectral weight is concentrated on a small number of eigenstates, this raises a practical question: can a favorable degree with respect to the success probability be selected in advance, without having to scan many degrees and retain the best one? This question is relevant whenever the degree is not fixed by the algorithm. The most direct application arises when the target accuracy only imposes a minimum degree $k\ge k_{\min}$, as is generic for polynomial methods whose degree is set by the spectral gap and the required precision~\cite{lin_tong_2020}, while the Chebyshev degree itself remains free above this threshold: since $p^{(T_k)}_{\mathrm{success}}$ varies strongly between neighboring degrees, a slightly deeper circuit can reduce the shot overhead substantially. For instance, for $N=8$ at $U/t=16$ (Fig.~\ref{fig:pk_vs_k}), increasing the degree from $k=2$ to $k=3$ raises $p^{(T_k)}_{\mathrm{success}}$ from $\approx0.23$ to $\approx0.94$, i.e., a fourfold reduction in shot overhead at the cost of a single additional walk step. A second, more specific application is the construction of a Krylov space from a noncontiguous set of degrees, where the degrees themselves are free parameters; we illustrate it in detail in Sec.~\ref{sssec:krylov}. In both cases, identifying favorable degrees without scanning them all, which would itself cost shots at every degree, is operationally useful. To this end, we propose below two inexpensive strategies to identify, a priori, Chebyshev degrees with a high success probability.

We quantify the effective number of eigenstates contributing to the input state through the participation ratio
\begin{equation}
    \mathrm{PR} = \frac{1}{\sum_n w_n^2}.
    \label{eq:participation_ratio}
\end{equation}
Recall from Eq.~\eqref{eq:spectral_decomposition} that $\sum_n w_n=1$ while $\sum_n w_n^2\le1$, so that $\mathrm{PR}\ge1$. The participation ratio is close to $1$ when the weight is concentrated on a few eigenstates, giving a few dominant frequencies and strong coherent oscillations in $p^{(T_k)}_{\mathrm{success}}$, and large when the weight is spread over the spectrum, in which case the oscillations average out over $k$.
Once the dominant pairs $\{w_n, \theta_n\}$ are known, Eq.~\eqref{eq:pk_spectral} determines $p^{(T_k)}_{\mathrm{success}}$ for every degree, allowing the optimal degree to be identified without executing the corresponding deep circuits. Writing $k^\star$ for the selected degree and $p_{\mathrm{med}}$ for the median of $p^{(T_k)}_{\mathrm{success}}$ over the considered degree window, we use $p^{(T_{k^\star})}_{\mathrm{success}}/p_{\mathrm{med}}$ as the figure of merit. The median represents the success probability expected from a degree chosen without spectral information, so the ratio measures the gain brought by the selection strategy. The largest achievable value of this ratio is $p_{\max}/p_{\mathrm{med}}$, with $p_{\max}=\max_k p^{(T_k)}_{\mathrm{success}}$ over the same window; we call $p_{\max}$ the oracle optimum, since reaching it requires knowing $p^{(T_k)}_{\mathrm{success}}$ at every degree. All results below use the window $1\le k\le200$ and the HF input state for $N=2,4,6,8$ and $U/t\in\{4,8,16\}$, i.e., $12$ cases.

\emph{The spectrum-informed strategy} keeps only the $m$ eigenstates with the largest weights $w_n$, whose index set we denote $\mathcal{M}_m$, and optimizes the resulting partial sum,
\begin{equation}
\label{eq:versionA}
    \begin{gathered}
    p^{(T_k)}_{\mathrm{success},\,m} = \sum_{n\in\mathcal{M}_m} w_n\cos^2(k\theta_n),\\
    k_A^\star = \arg\max_k \, p^{(T_k)}_{\mathrm{success},\,m}.
    \end{gathered}
\end{equation}
It requires the full eigendecomposition and therefore serves as a spectrum-informed benchmark rather than a practical spectrum-free protocol. With $m=5$, the selected degree lies within $5\%$ of the oracle optimum in $11$ of the $12$ cases and reaches on average $98\%$ of $p_{\max}$, showing that a small number of spectral components captures most of the relevant $k$-dependence.

\emph{The spectrum-free strategy} uses no spectral input. It requires $p^{(T_k)}_{\mathrm{success}}$ only at shallow degrees $k\leq k_\mathrm{fit}$, fits $m$ weights and angles $\{w_n^\mathrm{fit},\theta_n^\mathrm{fit}\}$ to those points by nonlinear least squares against the model in Eq.~\eqref{eq:pk_spectral}, and extrapolates the fitted sum to select
\begin{equation}
\label{eq:versionB}
    k_B^\star = \arg\max_k \, \sum_{n\le m}w_n^\mathrm{fit}\cos^2(k\,\theta_n^\mathrm{fit}) .
\end{equation}
Because $p^{(T_k)}_{\mathrm{success}}$ is a weighted sum of a small number of squared oscillatory components, a window of shallow circuits can constrain their frequencies and amplitudes, and the fitted model can predict a favorable deeper degree without classical diagonalization. We use $k_\mathrm{fit}=25$ and $m=5$, so that the $2m=10$ fitted parameters are fewer than the $k_\mathrm{fit}$ data points. In the results below, the shallow values of $p^{(T_k)}_{\mathrm{success}}$ are computed exactly; the effect of shot noise on the fit is not assessed here.

\begin{figure*}[tbp]
  \centering
  \includegraphics[width=0.94\textwidth]{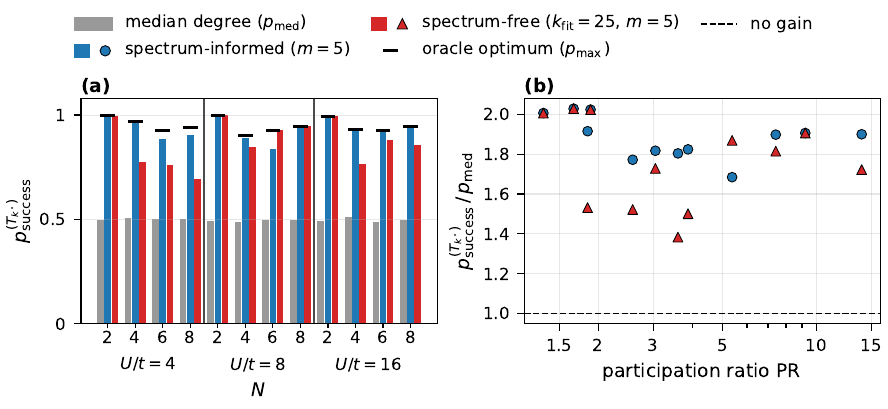}
  \caption{Degree-selection performance for HF input states at $N\leq8$ and $U/t\in\{4,8,16\}$, over the window $1\le k\le200$. \textbf{(a)} Success probability at the degree selected by the spectrum-informed strategy (blue, Eq.~\eqref{eq:versionA}, $m=5$) and the spectrum-free strategy (red, Eq.~\eqref{eq:versionB}, $k_\mathrm{fit}=25$, $m=5$), compared with the median-degree baseline $p_\mathrm{med}$ (grey) and the oracle optimum $p_{\max}$ (black horizontal lines). Vertical lines separate the three values of $U/t$. \textbf{(b)} Gain $p^{(T_{k^\star})}_{\mathrm{success}}/p_\mathrm{med}$ as a function of the participation ratio for the spectrum-informed (blue circles) and spectrum-free (red triangles) strategies; the dashed line marks no gain.}
  \label{fig:pick-k}
\end{figure*}

For the HF input state, the median-degree baseline is $p_{\mathrm{med}}\approx1/2$ (between $0.49$ and $0.51$) in all cases. The spectrum-informed strategy raises the per-circuit success probability to $98\%$ of the oracle optimum on average, corresponding to a gain of $1.7$--$2.0\times$, nearly independently of the participation ratio over the range studied ($\mathrm{PR}\approx1.4$--$14$). The spectrum-free strategy reaches $91\%$ of the oracle optimum on average, with gains of $1.4$--$2.0\times$, and never selects a degree below the median (Fig.~\ref{fig:pick-k}). Its performance depends on the length of the shallow window and on the number of fitted components. With $k_\mathrm{fit}=10$, which allows at most $m=4$ components, the gain ranges from $0.84\times$ to $2.0\times$: when nearby frequencies are not resolved, the extrapolation identifies the wrong revival and can select a degree below the median. The required number of components also grows with the fitting window, since a longer window resolves more of the beating between frequencies: at $k_\mathrm{fit}=25$, $m=3$ gives a minimum gain of only $0.46\times$, whereas $m=5$ gives $1.4\times$.

In the two settings identified at the beginning of this section, degree selection can therefore be useful. The tests above select the best degree over the full window $1\le k\le200$; restricting the search to $k\ge k_{\min}$, or ranking degrees to build a noncontiguous Krylov space (Sec.~\ref{sssec:krylov}), applies the same predicted $p^{(T_k)}_{\mathrm{success}}$ to a subwindow and was not tested separately. The spectrum-informed strategy reliably locates near-optimal degrees but requires the dominant eigenpairs. The spectrum-free strategy requires only shallow-degree data and, with $k_\mathrm{fit}=25$ and $m=5$, consistently improves on a blind choice for the HF states studied here; its main benefit is to avoid near-node degrees at which a full circuit execution would have a near-vanishing acceptance probability.

\subsubsection{Application to quantum Krylov method}
\label{sssec:krylov}
 
The success-probability analysis becomes operationally concrete when the qubitized walk is used as the primitive of a quantum Krylov method.
The Krylov subspace is spanned by the vectors
\begin{equation}
\label{eq:krylov_vectors}
    \ket{\varphi_k}=T_k(\widetilde H)\ket{\psi},\qquad k=0,\dots,D-1,
\end{equation}
and the ground-state energy is estimated from the regularized generalized eigenvalue problem $\mathcal{H}\mathbf{c}=\tilde E\,S\mathbf{c}$, with $S_{jk}=\braket{\varphi_j|\varphi_k}$ the Gram (overlap) matrix and $\mathcal{H}_{jk}=\bra{\varphi_j}\widetilde H\ket{\varphi_k}$ the Hamiltonian projected onto the $D$-dimensional Krylov subspace.

The Krylov vectors of Eq.~\eqref{eq:krylov_vectors} are not normalized and thus cannot be prepared directly as quantum states. Applying $\mathrm{PREP}^\dagger$ after $k$ walk steps and postselecting the ancilla register on $a=0$ would leave $\ket{\varphi_k}/\|\varphi_k\|$ in the system register with probability $p^{(T_k)}_{\mathrm{success}}$. 
Instead, Kirby et al.\ estimate only the Chebyshev moments~\cite{weisse_2006,Kirby_2023},
\begin{equation}
\label{eq:moments}
    m_k=\bra{\psi}T_k(\widetilde H)\ket{\psi},
\end{equation}
each as the expectation value of an observable, and $S$ and $\mathcal{H}$ are then assembled classically~\cite{Kirby_2023} as
\begin{equation}
\label{eq:krylov_matrices}
    \begin{aligned}
    S_{jk}&=\tfrac12\big(m_{j+k}+m_{|j-k|}\big),\\
    \mathcal{H}_{jk}&=\tfrac14\big(m_{j+k+1}+m_{|j+k-1|}+m_{|j-k+1|}+m_{|j-k-1|}\big).
    \end{aligned}
\end{equation}
A Krylov space of dimension $D$ therefore requires the moments up to $k_{\max}=2D-1$. 
In such a moment-based scheme no postselection is required. However, the success probability directly affects the resources (number of shots) needed to reach a given precision, as we now show.

The qubitization success probability of Eq.~\eqref{eq:pk_qubit} can be rewritten, using Eq.~\eqref{eq:krylov_matrices} with $m_0=1$, as
\begin{equation}
\label{eq:Skk}
    p^{(T_k)}_{\mathrm{success}}=\|\varphi_k\|^2=S_{kk}=\frac{1+m_{2k}}{2}.
\end{equation}
Degrees with small $p^{(T_k)}_{\mathrm{success}}$ produce Krylov vectors of small norm, whose moments are also small, $|m_k|=|\braket{\psi|\varphi_k}|\le\|\varphi_k\|$. As $m_{2k}$ is estimated from $N_{\mathrm{shots}}$ binary outcomes (Appendix~\ref{app:odd_readout}), its standard deviation is $\sqrt{(1-m_{2k}^2)/N_{\mathrm{shots}}}$, and that of $S_{kk}$ is $\sqrt{p^{(T_k)}_{\mathrm{success}}\big(1-p^{(T_k)}_{\mathrm{success}}\big)/N_{\mathrm{shots}}}$. Resolving the squared norm $S_{kk}$ of $\ket{\varphi_k}$ to a relative precision $\delta$ thus requires
\begin{equation}
\label{eq:nshots_norm}
    N_{\mathrm{shots}}\simeq\frac{1-p^{(T_k)}_{\mathrm{success}}}{\delta^{2}\,p^{(T_k)}_{\mathrm{success}}},
\end{equation}
so that the $1/p^{(T_k)}_{\mathrm{success}}$ overhead remains even without postselection: not as discarded shots, but as the shots needed to resolve a Krylov vector of small norm with a controlled precision.

The scheme of Kirby et al.~\cite{Kirby_2023} is based on moment measurements, each moment being estimated without a control qubit, by preparing the walk state $W^{\lfloor k/2\rfloor}\big(\ket{\mathrm{PREP}}\otimes\ket{\psi}\big)$ and measuring $R_\mathrm{PREP}$ for even $k$ and $\mathrm{SELECT}$ for odd $k$. 

Our scheme is also based on moment measurements, rather than on postselection as in the previous sections, which makes it directly comparable with that of Kirby et al.~\cite{Kirby_2023}.
The only difference is that we measure each moment with a Hadamard test~\cite{kitaev1995,aharonov_polynomial_2009}, as commonly done for overlaps in quantum subspace methods~\cite{parrish_mcmahon_2019,stair2019multireferencequantumkrylovalgorithm,klymko_2022,Cortes_QuantumKrylov}; the circuit is shown in Fig.~\ref{fig:hadamard_test} and its readout is derived in Appendix~\ref{app:odd_readout}.

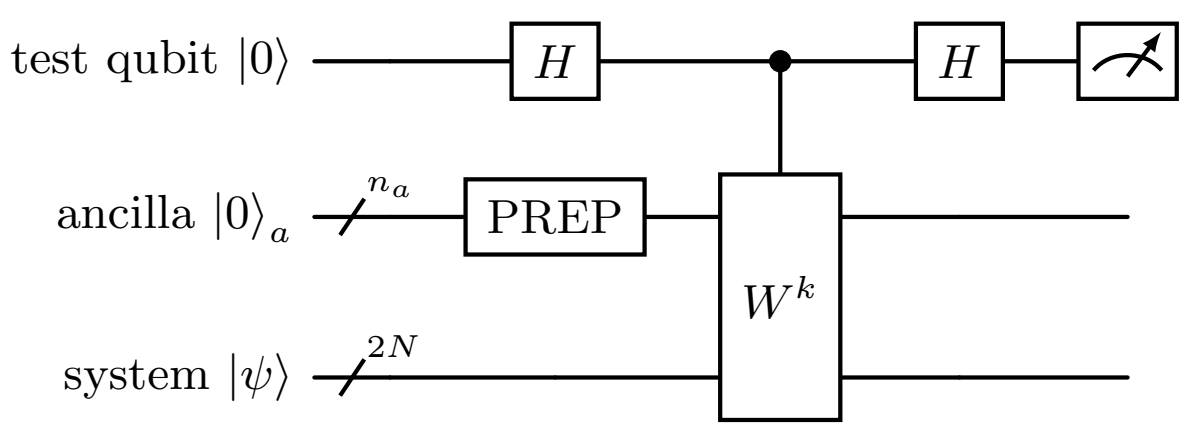
\begin{figure}[tbp]
\centering
\begin{quantikz}
\lstick{test qubit $\ket{0}$} & & \gate{H} & \ctrl{1} & \gate{H} & \meter{} \\
\lstick{ancilla $\ket{0}_a$} & \qwbundle{n_a} & \gate{\mathrm{PREP}} & \gate[2]{W^k} & & \\
\lstick{system $\ket{\psi}$} & \qwbundle{2N} & & & &
\end{quantikz}
\caption{Hadamard-test circuit for the Chebyshev moment $m_k=\bra{\psi}T_k(\widetilde H)\ket{\psi}$. The test qubit controls $k$ applications of the walk operator $W=R_\mathrm{PREP}\,\mathrm{SELECT}$ on $\ket{\mathrm{PREP}}\otimes\ket{\psi}$; only the test qubit is measured, and $m_k=(-1)^k(p_0-p_1)$ from the probabilities $p_{0,1}$ of its two outcomes, the sign $(-1)^k$ coming from the reflection convention of our implementation. The block-encoding ancillas are never measured, so no shot is discarded.}
\label{fig:hadamard_test}
\end{figure}
 
The two schemes offer different trade-offs between circuit depth and readout cost (Table~\ref{tab:readouts}). In terms of depth, the scheme of Kirby et al.\ requires only $\lfloor k/2\rfloor$ uncontrolled walk steps, i.e., a maximum depth of $D-1$ instead of $2D-1$ controlled steps for our Hadamard test scheme. 

In our Hadamard-test scheme, every shot contributes to the estimate of $m_k$, for even and odd $k$ alike, with standard deviation $\sqrt{(1-m_k^2)/N_{\mathrm{shots}}}\le1/\sqrt{N_{\mathrm{shots}}}$ and without any discard. We prefer this trade-off here. The number of shots is already set by the success probability through Eq.~\eqref{eq:nshots_norm}, which can be large for degrees with small $p^{(T_k)}_{\mathrm{success}}$. The term-by-term readout of the odd moments, which make up half of the moments required, never needs fewer shots than the Hadamard test for the same error, and needs at least $L-1$ times more when $q_k(i)$ is spread evenly over the $L$ terms (Appendix~\ref{app:odd_readout}, Eq.~\eqref{eq:odd_vs_hadamard}); this cost comes on top of Eq.~\eqref{eq:nshots_norm}. The Hadamard test avoids this additional cost at the price of a controlled walk and twice the circuit depth.

\begin{table*}[tbp]
\centering
\caption{Readouts of the Chebyshev moment $m_k$: number of walk steps, counted outcome, its probability, and number of shots carrying information, out of $N$ shots in total ($N_i$ for term $i$). The probabilities of the two readouts of Kirby et al.\ are derived in Appendix~\ref{app:odd_readout}.}
\label{tab:readouts}
\begin{tabular}{lcccc}
\toprule
Readout & Walk steps & Counted outcome & Probability & Useful shots \\
\midrule
Hadamard test & $k$ (controlled) & test qubit $=0$ & $(1+m_k)/2$ & all $N$ \\
Even $k$~\cite{Kirby_2023} & $k/2$ & $a=0$ after $\mathrm{PREP}^\dagger$ & $p^{(T_{k/2})}_{\mathrm{success}}=(1+m_k)/2$ & all $N$ \\
Odd $k$, term $i$~\cite{Kirby_2023} & $(k-1)/2$ & $a=i$ & $q_k(i)$, Eq.~\eqref{eq:odd_Pk} & $\approx N_i\,q_k(i)$ \\
\bottomrule
\end{tabular}
\end{table*}
 
\begin{figure*}[tbp]
    \centering
    \includegraphics[width=0.71\textwidth]{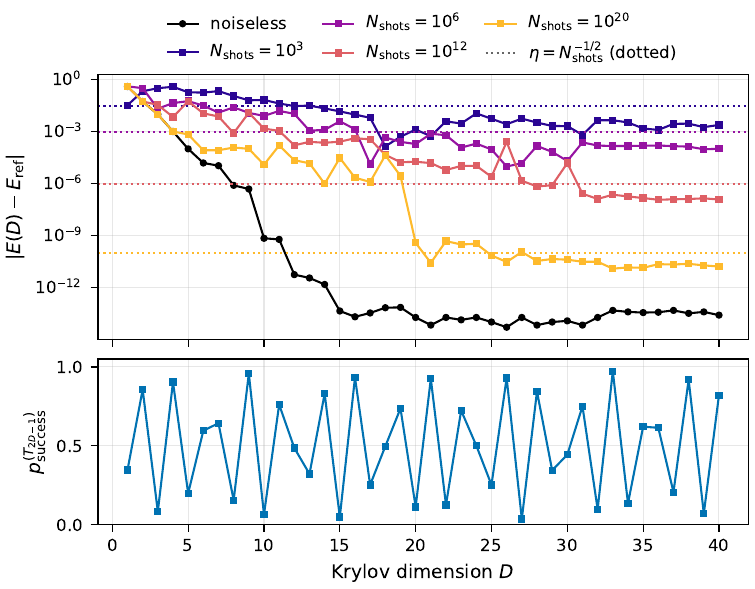}
    \caption{Finite-shot convergence and success-probability variation in the quantum Krylov calculation for the $N=4$, $U/t=2$ Hubbard model with a HF input state. \textbf{Upper panel}: Ground-state energy error versus Krylov dimension $D$ for $N_{\mathrm{shots}}=10^3$ and $10^6$ (sampled Hadamard-test circuits) and $10^{12}$ and $10^{20}$ (Gaussian moment noise $\eta=N_{\mathrm{shots}}^{-1/2}$), together with the noiseless curve.    Dotted lines mark $\eta=N_\mathrm{shots}^{-1/2}$, in the color of the corresponding curve. \textbf{Lower panel}: Success probability $p^{(T_{2D-1})}_{\mathrm{success}}$ of the highest-degree moment required at each Krylov dimension.}
    \label{fig:lanczos_combined}
\end{figure*}

\begin{figure*}[tbp]
    \centering
    \includegraphics[width=0.84\textwidth]{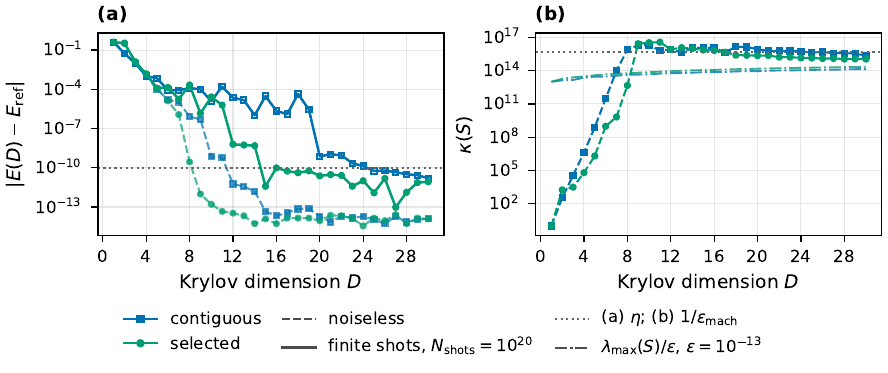}
    \caption{A priori construction of the Krylov space from degrees with high success probability ($p^{(T_k)}_{\mathrm{success}} \geq 1/2$), for $N=4$, $U/t=2$, compared with the standard contiguous construction. \textbf{(a)} Ground-state energy error $|E(D)-E_\mathrm{ref}|$ versus Krylov dimension $D$, where $E(D)$ is the Krylov estimate with $D$ basis vectors and $E_\mathrm{ref}$ the exact ground-state energy, for the contiguous (blue) and selected-degree (green) constructions, noiseless (dashed) and with finite-shot noise for $N_\mathrm{shots}=10^{20}$, emulated by Gaussian noise of standard deviation $\eta=N_\mathrm{shots}^{-1/2}$ on the moments (solid); the dotted line marks $\eta$. \textbf{(b)} Condition number $\kappa(S)=\lambda_{\max}/\lambda_{\min}$ of the overlap matrix $S$ versus $D$ for each construction (noiseless). The dash-dotted lines, nearly coincident for the two constructions, mark $\lambda_{\max}(S)/\varepsilon$ with $\varepsilon=10^{-13}$: above them, the smallest eigenvalue of $S$ falls below the threshold and the corresponding direction is discarded. The dotted line marks the inverse of the double-precision machine accuracy, $\approx4.5\times10^{15}$, beyond which the smallest eigenvalue of $S$ is at the level of round-off.}
    \label{fig:selected_krylov}
\end{figure*}

The results of Figs.~\ref{fig:lanczos_combined} and~\ref{fig:selected_krylov} use four shot budgets per moment. For $N_{\mathrm{shots}}=10^3$ and $10^6$, the Hadamard-test circuits are sampled. The larger budgets, $N_{\mathrm{shots}}=10^{12}$ and $10^{20}$, are too costly to sample in simulation; for these, the exact moments are perturbed by Gaussian noise of standard deviation $\eta=N_{\mathrm{shots}}^{-1/2}$. In the regularized solution of the generalized eigenvalue problem, directions of $S$ with eigenvalue below a threshold $\varepsilon$ are discarded~\cite{Kirby_2023,Epperly_2022,Lee_2024}. Without noise we use a fixed threshold $\varepsilon=10^{-13}$. With noise, $\varepsilon$ is selected for each $D$ among $40$ logarithmically spaced values between $10^{-14}$ and $10^{-1}$, restricted to $\varepsilon\ge10\,\eta$, as the smallest value within the widest plateau of the extracted energy, which makes the result insensitive to the exact cut. Krylov vectors with $p^{(T_k)}_{\mathrm{success}}\lesssim\varepsilon$ therefore cannot be resolved and contribute to the ill-conditioning of the overlap matrix. For the present calculation, the energy error decreases rapidly with Krylov dimension until statistical uncertainty becomes dominant, after which the convergence reaches a shot-limited plateau. The plateau level tracks the nominal moment-noise scale $\eta=1/\sqrt{N_{\mathrm{shots}}}$, and increasing the shot budget lowers this plateau, as shown in the upper panel of Fig.~\ref{fig:lanczos_combined}.
 
This motivates asking whether the Krylov space can be constructed using degrees with sufficiently large success probability. In Fig.~\ref{fig:selected_krylov}, we therefore construct the Krylov space a priori from degrees satisfying $p^{(T_k)}_{\mathrm{success}} \geq 1/2$. The selected-degree construction keeps the condition number $\kappa(S)$ below that of the standard contiguous construction for $3\le D\le8$, before both reach the round-off limit, and the improvement in convergence is already visible in the noiseless calculation. The corresponding finite-shot calculations retain the same trend. This improvement comes at the cost of a larger maximum Chebyshev degree and therefore deeper circuits. Thus, the success probability is not only a shot-overhead indicator for individual moments; it also provides a direct diagnostic for identifying Chebyshev degrees whose corresponding Krylov vectors may contribute to poor conditioning.

\subsection{QSVT success probability}
\label{ssec:qsvt}

QSVT generalizes the qubitized walk from a single Chebyshev polynomial to a broad class of polynomial transformations $P(\widetilde H)$ satisfying the required parity and boundedness conditions~\cite{qsvt,martyn2021}, implemented by the phase sequence of Eqs.~\eqref{eq:qsvt_equation} and~\eqref{eq:qsvt_block}. QSVT introduces a qualitatively different normalization problem: unlike LCU and qubitization block-encodings, where the scaling is inherited from the LCU $1$-norm, the polynomial transformation introduces an additional scaling factor that directly affects both the success probability and the feasibility of phase synthesis.

We write the target polynomial, before scaling, in the Chebyshev basis,
\begin{equation}
\label{eq:dev_P}
    P(x)=\sum_{k=0}^{d} a_k\,T_k(x),
\end{equation}
where $d$ is its degree ($a_d\neq0$) and the real coefficients $a_k$ vanish whenever $k$ and $d$ have different parity, so that $P$ has the definite parity of $d$. Let $C>0$ denote the scaling factor used to obtain the QSVT-feasible polynomial $f(x)=P(x)/C$. The resulting success probability is
\begin{equation}
\label{eq:pqsvt}
    p^{(P/C)}_{\mathrm{success}}(\ket{\psi}) = \frac{\|P(\widetilde H)\ket{\psi}\|^2}{C^2} = \frac{1}{C^2}\sum_n w_n\,|P(\tilde{E}_n)|^2.
\end{equation}
For a given target polynomial and input state, the numerator is fixed, so the scaling factor $C$ directly controls the QSVT success probability. It is constrained by the feasibility condition $\sup_{x\in[-1,1]}|P(x)/C|\le1$~\cite{qsvt,martyn2021}, which requires
\begin{equation}
    C \geq C_\mathrm{opt}=\sup_{x\in[-1,1]}|P(x)|.
\end{equation}
For a fixed target polynomial, the success probability is therefore maximized by choosing $C$ as close as possible to $C_\mathrm{opt}$ from above, so that feasibility is maintained. Estimating or certifying $C_\mathrm{opt}$ is thus central to the analysis below. Unless stated otherwise, the target polynomials are randomly generated in the Chebyshev basis with definite parity.

\subsubsection{Choice of the polynomial scaling factor}
\label{sssec:csub}

Several constants bound $C_\mathrm{opt}$ from above, and they form an ordered chain
\begin{equation}
\label{eq:c_chain}
    \begin{split}
    C_\mathrm{grid} &\;\lesssim\; C_\mathrm{dense} \;\lesssim\; C_\mathrm{num} \approx C_\mathrm{opt}\\
    &\;\le\; C_\mathrm{Markov} \;\le\; \Lambda \;\le\; \|\mathbf a\|_1 ,
    \end{split}
\end{equation}
where $C_\mathrm{grid}$ and $C_\mathrm{dense}$ are the maxima of $|P|$ sampled on a coarse (4000 points) and a fine (200000 points) grid, $C_\mathrm{num}$ denotes a numerical evaluation of $C_\mathrm{opt}$, obtained by solving $P'(x) = 0$ and evaluating $P$ at the resulting stationary points and the interval endpoints, $C_\mathrm{Markov}$ is a certified bound derived below, $\Lambda=\max_{|z|=1}|\sum_k a_k z^k|$ is the sup norm of the coefficient series on the unit circle, and $\|\mathbf a\|_1=\sum_k|a_k|$ is the coefficient $\ell_1$-norm. The inequalities involving $C_\mathrm{grid}$ and $C_\mathrm{dense}$ are empirical properties of the sampled grids, not certified bounds: they provide lower estimates of the supremum because they only evaluate $P$ at discrete points. Consequently, normalizing with an undershooting grid estimate can produce a polynomial that exceeds unit magnitude between grid points. The three quantities to the right of $C_\mathrm{opt}$ provide valid upper bounds.

The maximum of $|P|$ on the closed interval $[-1,1]$ is attained either at a critical point, where $P'(x)=0$, or at an endpoint $x=\pm1$. 
Using Eq.~\eqref{eq:dev_P} and $T_k'=k\,U_{k-1}$ with $U$ the second-kind Chebyshev polynomials, the derivative is
\begin{equation}
\label{eq:pprime}
    P'(x)=\sum_{k=1}^{d} a_k\,k\,U_{k-1}(x) .
\end{equation}
The stationary points of $P$ in $[-1,1]$ are obtained numerically as the real roots of $P'$, computed as the eigenvalues of its companion matrix. Evaluating $|P(x)|$ at these points and at the endpoints $x=\pm1$ gives the numerical estimate $C_\mathrm{num}$ of $\sup_{x\in[-1,1]}|P(x)|$ without requiring a dense grid.
Unlike the Markov construction below, this critical-point procedure is not a certified upper bound.

Normalization by the infinity norm is standard in QSVT, and the numerical practice is to estimate it by grid sampling. A grid maximum undershoots the true supremum, however, and so gives no guarantee of the feasibility condition $\sup|P/C|\le1$. The grid estimate can be corrected upward into a certified bound by accounting for the largest rise the polynomial can hide between two nodes. Let $x^\star$ be the true maximizer, $|P(x^\star)|=C_\mathrm{opt}$. With nodes spaced $h$ apart, $x^\star$ lies within $h/2$ of some node $x_i$, and by the mean value theorem~\cite{meanvalue}, there is a $\xi$ between $x_i$ and $x^\star$ with
\begin{equation}
\label{eq:mvt_step}
    \begin{split}
    C_\mathrm{opt}=|P(x^\star)|
    &\le |P(x_i)|+|P'(\xi)|\,|x^\star-x_i|\\
    &\le C_\mathrm{grid}+\frac{h}{2}\,\max_{x\in[-1,1]}|P'(x)| ,
    \end{split}
\end{equation}
where we used the triangle inequality, $|P(x_i)|\le C_\mathrm{grid}$, which holds because $x_i$ is itself one of the nodes over which $C_\mathrm{grid}=\max_i|P(x_i)|$ is taken (the grid includes both endpoints $x=\pm1$), and $|x^\star-x_i|\le h/2$. The derivative is bounded by the Markov inequality~\cite{markov_inequality,borwein_erdelyi1995}, which holds for any polynomial of degree $d$ on $[-1,1]$, with $d$ the degree of the target polynomial $P$ of Eq.~\eqref{eq:dev_P}:
\begin{equation}
    \max_{x\in[-1,1]}|P'(x)|\le d^2\,\max_{x\in[-1,1]}|P(x)|=d^2\,C_\mathrm{opt}.
\end{equation}
Substituting into Eq.~\eqref{eq:mvt_step} and solving for $C_\mathrm{opt}$ gives
\begin{equation}
\label{eq:c_markov}
    C_\mathrm{opt}\le C_\mathrm{Markov}\equiv\frac{C_\mathrm{grid}}{1-\tfrac{h}{2}d^2},
\end{equation}
valid whenever $hd^2/2<1$, which requires a grid sufficiently fine relative to the polynomial degree. Unlike $C_\mathrm{num}$, Eq.~\eqref{eq:c_markov} provides a provable upper bound and therefore guarantees feasibility.

\begin{figure*}[tbp]
    \centering
    \includegraphics[width=0.84\textwidth]{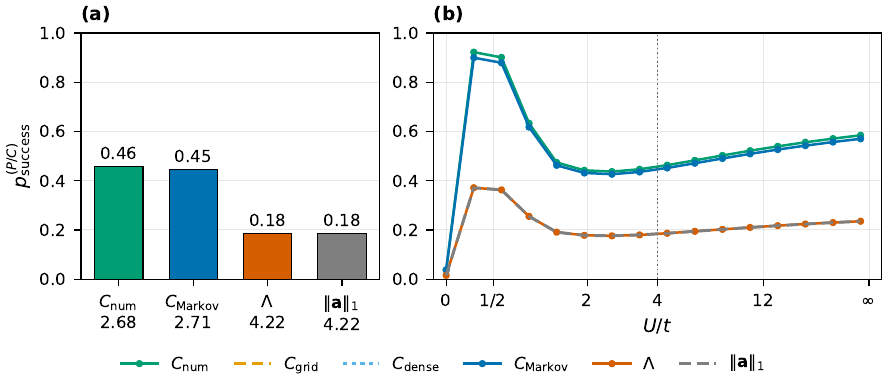}
    \caption{QSVT success probability and the cost of the normalization constant, for the fixed degree-7 target $P=-2.37\,T_1+1.13\,T_3+0.72\,T_7$ and an $N=2$ HF input state. \textbf{(a)} Success probability at $U=4t$ for each normalization choice, with the value of $C$ under each label. The unscaled numerator is $\|P(\widetilde H)\ket{\psi_\mathrm{HF}}\|^2=3.278$; $C_\mathrm{num}$, $C_\mathrm{Markov}$ and $\Lambda$ (equal to $\|\mathbf a\|_1$ for this target) give $p^{(P/C)}_{\mathrm{success}}\approx0.46$, $0.45$ and $0.18$, respectively. \textbf{(b)} Success probability as a function of $U/t$ (horizontal axis scaled as $U/(U+4t)$; dotted line: $U=4t$, the value used in (a)). The curves for $C_\mathrm{grid}$ (dashed) and $C_\mathrm{dense}$ (dotted) coincide with $C_\mathrm{num}$, and the curve for $\|\mathbf a\|_1$ (dashed) coincides with $\Lambda$. The tight estimates $C_\mathrm{num}$ and $C_\mathrm{Markov}$ form the upper group, while the looser bounds $\Lambda$ and $\|\mathbf a\|_1$ give success probabilities smaller by a factor $\approx2.5$ across the $U/t$ range studied.}
    \label{fig:qsvt_normalization}
\end{figure*}

Since the numerator of Eq.~\eqref{eq:pqsvt} is fixed by the target, the choice among the constants of Eq.~\eqref{eq:c_chain} directly sets $p^{(P/C)}_{\mathrm{success}}$, as illustrated in Fig.~\ref{fig:qsvt_normalization}(a) for a fixed degree-7 target. The certified Markov bound $C_\mathrm{Markov}=2.71$ gives $p^{(P/C)}_{\mathrm{success}}\approx0.45$, only slightly below the tight value $\approx0.46$, whereas the loose bound $\Lambda=4.22$ reduces the success probability to $\approx0.18$. Therefore, for this example, a looser normalization alone produces a $(\Lambda/C_\mathrm{num})^2\approx2.5\times$ increase in the shot overhead. This ordering persists across the correlation-strength sweep considered here: Fig.~\ref{fig:qsvt_normalization}(b) shows the same separation between $C_\mathrm{num}$, $C_\mathrm{Markov}$, and the looser $\Lambda$ and $\|\mathbf a\|_1$ choices across the full $U/t$ range studied. The gap between the tight and loose constants is expected to widen with the polynomial degree. $\Lambda$ and $\|\mathbf a\|_1$ depend only on the Chebyshev coefficients of the target and accumulate their magnitudes, e.g., $\|\mathbf a\|_1=d+1$ for constant coefficients, whereas $C_\mathrm{opt}=\sup_{[-1,1]}|P|$ can remain much smaller owing to cancellations between Chebyshev terms of different order at each $x$ (Sec.~\ref{sssec:qsvt_degree}). In contrast, the classical cost of a tight estimate is independent of the system size: $C_\mathrm{num}$ requires the roots of $P'$, i.e., the eigenvalues of a $(d-1)\times(d-1)$ companion matrix, at cost $O(d^3)$, and $C_\mathrm{Markov}$ requires evaluating $P$ on a grid with $h<2/d^2$, i.e., $O(d^2)$ nodes at $O(d)$ cost each, hence $O(d^3)$. This one-time classical cost is negligible compared with the shot overhead $(C/C_\mathrm{opt})^2$ incurred at every execution of the circuit with a loose normalization, so the certified bound $C_\mathrm{Markov}$ should be preferred in practice.

\subsubsection{Effect on angle finding}
\label{sssec:angles}

The choice of $C$ represents a success probability trade-off, and it also determines whether phase finding is feasible and how robustly the numerical solver converges. The QSVT phase factors implementing the scaled operator $P/C$ are found by the Newton iteration of the \texttt{sym\_qsp} solver of \texttt{pyqsp}, following the phase-factor evaluation approach of Refs.~\cite{pyqsp,symqsp}, whose numerical conditioning depends on the distance between the normalized target $P/C$ and the feasibility boundary $\sup|P/C|=1$. Three regimes emerge, separated by the sign and size of the margin $C-C_\mathrm{opt}$.

\begin{figure}[tbp]
    \centering
    \includegraphics[width=\linewidth]{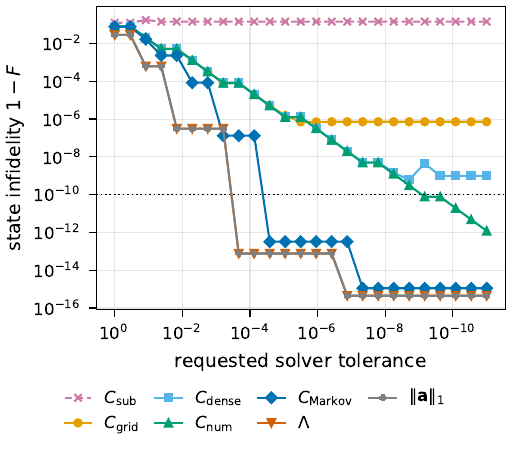}
    \caption{State infidelity $1-F$ of the realized QSVT circuit versus the tolerance requested from the \texttt{pyqsp} phase solver, for the normalization constants $C_\mathrm{sub}=0.75$, $C_\mathrm{grid}\approx C_\mathrm{dense}\approx C_\mathrm{num}=0.79$, $C_\mathrm{Markov}=0.795$ and $\Lambda=\|\mathbf a\|_1=0.898$ (the last two curves coincide). Degree-5 target $P=0.13\,T_1-0.13\,T_3+0.64\,T_5$, $N=2$, $U/t=4$, HF input state. The dashed curve ($C_\mathrm{sub}$) is infeasible. The horizontal dotted line marks the acceptance threshold $1-F=10^{-10}$ on the state infidelity.}
    \label{fig:qsvt_wrongphases}
\end{figure}

When $C<C_\mathrm{opt}$, the scaled polynomial violates the necessary boundedness condition $\sup_{[-1,1]}|P/C| \leq 1$, so it is not a feasible QSVT target on the full interval. In the numerical solver, this manifests as failure to converge to the required residual; we therefore reject phase sequences whose residual exceeds $10^{-9}$. This is the case for $C_\mathrm{sub}=0.95\,C_\mathrm{dense}$, about $5\%$ below the supremum: phase finding does not converge at any tested degree ($d=2$--$8$, Table~\ref{tab:qsvt_angles}), and for the degree-5 target the realized state keeps an infidelity $1-F\sim10^{-1}$ whatever the solver tolerance (Fig.~\ref{fig:qsvt_wrongphases}).
When $C$ is obtained from a grid estimate and chosen with only a grid-scale margin relative to $C_\mathrm{opt}$, phase finding can be fragile. The grid estimates $C_\mathrm{grid}$ and $C_\mathrm{dense}$ are not certified: although they often lie extremely close to the true supremum, they can undershoot it when a polynomial maximum falls between sampled nodes. In such cases, the nominal margin $C-C_\mathrm{opt}$ becomes negative and the solver can stall or converge to an inaccurate transformation.

If the margin is small but positive, phase finding is both fast and accurate. The certified $C_\mathrm{Markov}$ clears the supremum by about one percent and converges in $7$--$9$ Newton iterations to a residual of order $10^{-13}$ or below at every degree tested; the numerical estimate $C_\mathrm{num}$ converges in a fixed $20$ iterations to $\sim10^{-13}$.
Figure~\ref{fig:qsvt_wrongphases} shows how the choice of $C$ sets the precision of the realized state as the tolerance requested from the phase solver is tightened. Down to a tolerance of $\sim10^{-5}$, $C_\mathrm{grid}$, $C_\mathrm{dense}$ and $C_\mathrm{num}$, which all round to $0.79$, give the same state infidelity. Beyond this point the two grid estimates saturate, at $1-F\sim10^{-6}$ for $C_\mathrm{grid}$ and $\sim10^{-9}$ for $C_\mathrm{dense}$, consistent with a slight undershoot of $C_\mathrm{opt}$, whereas $C_\mathrm{num}$ keeps improving and reaches $\sim10^{-12}$ at a tolerance of $10^{-11}$. The margined constants reach a higher precision at much looser tolerances: $1-F\lesssim10^{-12}$ once the tolerance reaches $10^{-4}$--$10^{-5}$, and machine precision ($\sim10^{-14}$) from $\sim10^{-7}$, for $C_\mathrm{Markov}$, $\Lambda$ and $\|\mathbf a\|_1$. With an acceptance threshold of $1-F=10^{-10}$ on the state infidelity, the two grid estimates never qualify but $C_\mathrm{Markov}$ already qualifies from $\sim10^{-5}$. A smaller scaling factor therefore gives a higher success probability at the cost of precision on the phases, and hence on the prepared state. $C_\mathrm{Markov}$ resolves this trade-off favorably: for this target, its success probability is only a factor $(C_\mathrm{Markov}/C_\mathrm{num})^2\approx1.01$ below that of $C_\mathrm{num}$, against $\approx1.3$ for $\Lambda$ and $\|\mathbf a\|_1$, while it reaches the same machine-precision fidelity as the looser bounds. For the targets studied here, $C_\mathrm{Markov}$ is therefore the operationally preferred choice: it is the tightest certified bound considered, preserving a high success probability while providing a small positive feasibility margin that ensures both the convergence of phase finding and an accurate state preparation.
This result applies to the compilation of any QSVT circuit, since normalizing the target polynomial is a compilation step common to all QSVT applications. The Markov certificate turns a cheap grid estimate into a provable bound that guarantees the compiled phase sequence implements the intended polynomial, at a cost of a few percent in success probability for the targets studied here, while reaching machine-precision fidelity at solver tolerances several orders of magnitude looser than the tight numerical estimate.

\begin{table*}[tbp]
    \centering
    \caption{Phase-finding iterations and final residual errors of \texttt{sym\_qsp} as a function of target degree \(d\) for the normalization constants considered (same targets as Fig.~\ref{fig:qsvt_wrongphases} for \(d=5\)). The infeasible \(C_{\rm sub}=0.95\,C_{\rm dense}\) never reaches the required residual. The grid-based normalization fails to reach it at \(d=3\), \(5\) and \(8\), reflecting its lack of a certified margin above \(C_{\rm opt}\). The certified \(C_{\rm Markov}\) converges at all tested degrees with a small positive margin.}
    \label{tab:qsvt_angles}
    \renewcommand{\arraystretch}{1.15}
    \begin{tabular}{r cc cc cc cc}
    \toprule
    & \multicolumn{2}{c}{$C_\mathrm{sub}$} & \multicolumn{2}{c}{$C_\mathrm{grid}$} & \multicolumn{2}{c}{$C_\mathrm{num}$} & \multicolumn{2}{c}{$C_\mathrm{Markov}$} \\
    \cmidrule(lr){2-3}\cmidrule(lr){4-5}\cmidrule(lr){6-7}\cmidrule(lr){8-9}
    $d$ & iter & residual & iter & residual & iter & residual & iter & residual \\
    \midrule
    2 & -- & $\mathbf{5.9\times10^{-2}}$ & 20 & $5.5\times10^{-13}$ & 20 & $5.5\times10^{-13}$ & 9 & $5.6\times10^{-17}$ \\
    3 & -- & $\mathbf{1.1\times10^{-1}}$ & -- & $\mathbf{1.3\times10^{-7}}$ & 20 & $7.8\times10^{-13}$ & 8 & $8.9\times10^{-16}$ \\
    4 & -- & $\mathbf{7.9\times10^{-2}}$ & 20 & $5.5\times10^{-13}$ & 20 & $5.5\times10^{-13}$ & 8 & $2.5\times10^{-16}$ \\
    5 & -- & $\mathbf{1.0\times10^{-1}}$ & -- & $\mathbf{5.1\times10^{-6}}$ & 20 & $7.3\times10^{-13}$ & 7 & $2.0\times10^{-13}$ \\
    6 & -- & $\mathbf{7.6\times10^{-1}}$ & 20 & $6.1\times10^{-13}$ & 20 & $6.1\times10^{-13}$ & 7 & $8.9\times10^{-16}$ \\
    7 & -- & $\mathbf{2.3}$ & 20 & $5.5\times10^{-13}$ & 20 & $5.5\times10^{-13}$ & 7 & $1.9\times10^{-16}$ \\
    8 & -- & $\mathbf{1.5\times10^{-1}}$ & -- & $\mathbf{7.9\times10^{-7}}$ & 20 & $5.7\times10^{-13}$ & 7 & $3.2\times10^{-16}$ \\
    \bottomrule
    \end{tabular}
\end{table*}

\subsubsection{Dependence on polynomial degree and coefficient growth}
\label{sssec:qsvt_degree}
\begin{figure*}[tbp]
    \centering
    \includegraphics[width=0.84\textwidth]{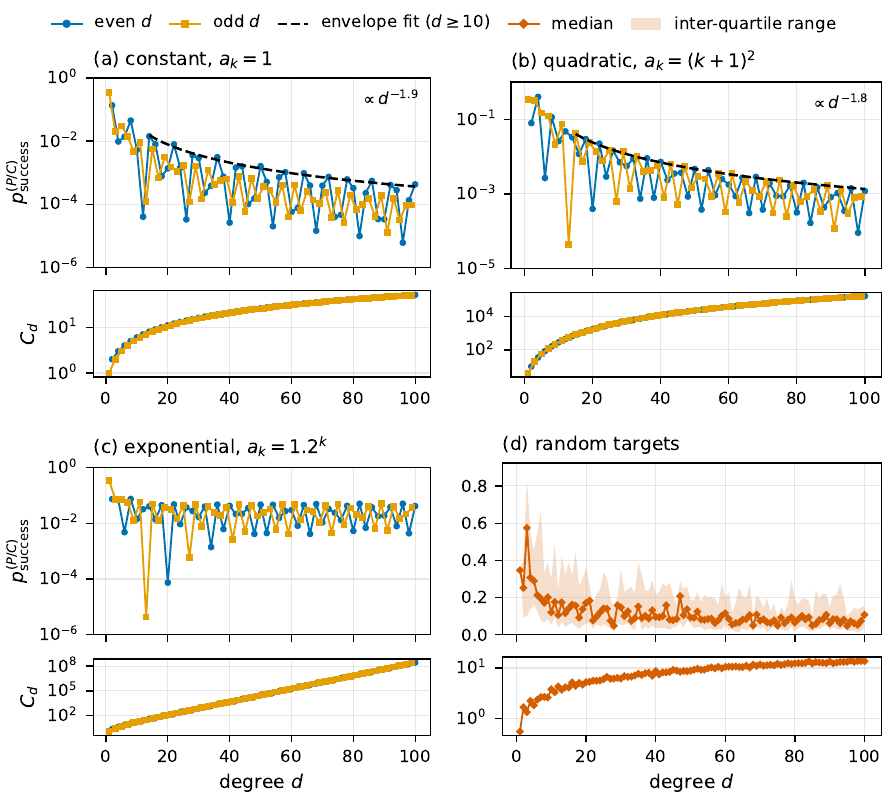}
    \caption{QSVT success probability versus polynomial degree $d$ for four coefficient profiles, $N=2$, $U=4t$, HF input state. Upper panels: $p^{(P/C)}_{\mathrm{success}}$, split by the parity of $d$; lower panels: normalization $C_d=\sup_{[-1,1]}|P_d|$. All targets have the parity of $d$, $P_d=\sum_j a_{d-2j}T_{d-2j}$. \textbf{(a)} Constant ($a_k=1$) and \textbf{(b)} quadratically growing ($a_k=(k+1)^2$) coefficients: the upper envelope decays as $d^{-\gamma}$, with $\gamma\approx1.9$ and $1.8$ (dashed lines, fit for $d\ge10$). \textbf{(c)} Exponential coefficients ($a_k=1.2^k$): $C_d$ grows by about 8 orders of magnitude, while $p^{(P/C)}_{\mathrm{success}}$ oscillates without decaying. \textbf{(d)} Random targets: median over 40 draws (line) and interquartile range (band).}
    \label{fig:qsvt_degree_profiles}
\end{figure*}
The QSVT success probability depends on the polynomial degree through the interplay between the normalization $C_d=\sup_{x\in[-1,1]}|P_d(x)|$, which enters as $1/C_d^2$, and the spectral response $\|P_d(\widetilde H)\ket{\psi}\|^2$. We index the target by its degree $d$, since we study how the success probability depends on $d$, and write Eq.~\eqref{eq:dev_P} as
\begin{equation}
    P_d(x)=\sum_{j=0}^{\lfloor d/2\rfloor} a_{d-2j}\,T_{d-2j}(x).
\end{equation}
The way $p^{(P/C)}_{\mathrm{success}}$ scales with $d$ is then governed by how fast the coefficients $a_k$ grow. In Krylov-type applications, the Lanczos polynomials expressed in the Chebyshev basis have coefficients that grow with degree~\cite{oqkd}; we therefore consider polynomial and exponential growth laws to bracket this behavior.
Consider first coefficients that are constant or grow polynomially, $a_k=(k+1)^\alpha$ with $\alpha\ge0$. Two sums then control the success probability. Since $a_k\ge0$ and $|T_k(x)|\le1$, the normalization is reached at $x=1$, where $T_k(1)=1$ for every $k$, so that all terms of $P_d$ add coherently: $C_d=\sum_k a_k=O(d^{\alpha+1})$. For a spectral weight spread uniformly over the angles $\theta_n$, the Chebyshev polynomials are orthogonal on average~\cite{mason_chebyshev_2002}, so the cross terms of $P_d^2$ cancel and only the squared coefficients remain: $\langle P_d^2\rangle=a_0^2+(1/2)\sum_{k\ge1}a_k^2=O(d^{2\alpha+1})$. The $d$ terms of $P_d$ thus add coherently in the normalization but incoherently in the numerator, so that
\begin{equation}
\label{eq:qsvt_dinv}
    p^{(P/C)}_{\mathrm{success},\,\mathrm{flat}}(d)=\frac{\langle P_d^2\rangle}{C_d^2}=O(d^{-1}),
\end{equation}
independently of $\alpha$: polynomial coefficient growth changes the normalization but not the $d^{-1}$ envelope of the success probability. For the concentrated HF input state the decay is faster than this flat-spectrum estimate: in Figs.~\ref{fig:qsvt_degree_profiles}(a) and~\ref{fig:qsvt_degree_profiles}(b) ($N=2$, $U=4t$), a power-law fit for $d\ge10$ shows that the upper envelope of $p^{(P/C)}_{\mathrm{success}}$ decays as $d^{-\gamma}$, with $\gamma\approx1.9$ for constant coefficients and $\gamma\approx1.8$ for both linear (not shown) and quadratic growth. The exponent thus depends only weakly on the coefficient profile, indicating that the decay is set by the input-state spectrum.
For exponentially growing coefficients, $a_k=r^k$ with $r>1$, the highest-degree terms dominate both the normalization, $C_d=\Theta(r^d)$, and the polynomial itself: the normalized polynomial $P_d/C_d$ is a combination of its last few Chebyshev terms, with weights decaying as $r^{-2j}$. As a result, the success probability oscillates with $d$ around an approximately constant level, of order $10^{-2}$ for $r=1.2$ (Fig.~\ref{fig:qsvt_degree_profiles}(c)), even though $C_d$ grows by about 8 orders of magnitude. Exponential coefficient growth alone thus does not drive the success probability to zero.
For generic random coefficient profiles, both the normalization and the spectral response fluctuate between realizations. The median success probability over draws decreases over the first $\sim20$ degrees and then remains nearly constant at $\approx0.1$ (Fig.~\ref{fig:qsvt_degree_profiles}(d)).

These observations bear on Krylov and Lanczos-based transformations, where high-degree polynomials can entail rapidly growing normalization~\cite{oqkd}. A large normalization alone does not determine the success probability: the spectral response of the polynomial on the input-state spectrum matters equally. The precise interplay of coefficient growth, spectral structure, and Krylov convergence is problem dependent, so the success probability should be monitored alongside the Krylov dimension and polynomial degree.

\subsubsection{Dependence on system size}
\label{ssec:qsvt_size}
\begin{figure*}[tbp]
    \centering
    \includegraphics[width=0.84\textwidth]{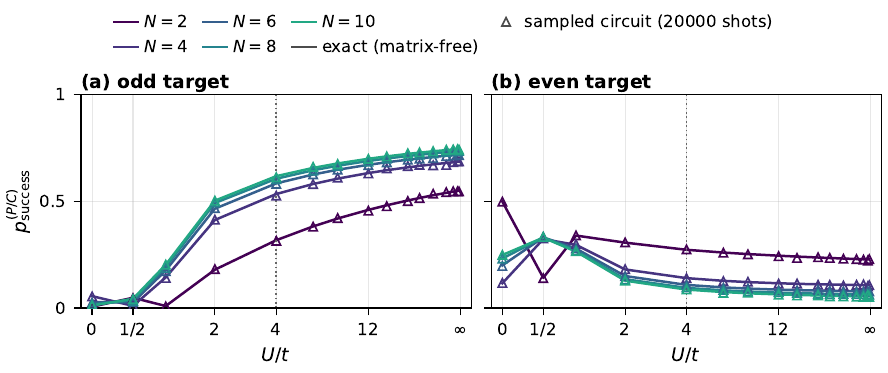}
    \caption{System-size and interaction-strength dependence of the QSVT success probability for fixed target polynomials of odd and even parity, HF input state, $N=2$--$10$ (color as in Fig.~\ref{fig:combined}(b)). \textbf{(a)} Odd-degree target $P(x)=0.10\,T_1-0.54\,T_3+0.36\,T_5$. \textbf{(b)} Even-degree target $P(x)=0.13\,T_0-0.13\,T_2+0.64\,T_4$; the isolated low value of the $N=2$ curve at $U=t/2$ is discussed in the text. Lines show exact (matrix-free) values computed without sampling; triangles show sampled-circuit estimates ($N_\mathrm{shots}=20000$), which agree with them to within $0.01$. The horizontal axis is scaled as $U/(U+4t)$. The two target families exhibit opposite $U/t$ trends.}
    \label{fig:combined_qsvt}
\end{figure*}
The system-size dependence differs between the even and odd target families. Figures~\ref{fig:combined_qsvt}(a) and~\ref{fig:combined_qsvt}(b) show the success probability across $U/t$ for system sizes $N=2$ to $10$ at fixed odd- and even-degree targets. For the odd target the success probability rises with $U/t$ and orders regularly with system size, the curves bunching for $N\ge4$; for the even target it falls with $U/t$ and the $N=2$ curve sits above the rest. As for LCU (Sec.~\ref{sec:lcu}), the largest finite-size changes occur between the smallest systems, with no systematic decay of $p^{(P/C)}_{\mathrm{success}}$ for $N\ge4$. At small $U/t$, the HF state is close to the ground state (participation ratio $\approx1$), so that $p^{(P/C)}_{\mathrm{success}}\approx|P(\tilde E_0)|^2/C^2$ is set by the position of the normalized ground-state energy relative to the zeros of $P$. At $U=0$, $\tilde E_0=-\Omega_s/\lambda_s=-\sqrt{p_\mathrm{bound}}$, which equals $-1$ for $N=2$ but $\approx-0.85$ for $N=4$ (Sec.~\ref{sec:lcu}). As $U$ increases, $\tilde E_0$ moves away from $-1$; for $N=2$ it crosses the zero of the even target at $x\approx-0.92$ between $U=0$ and $U=t/2$, which produces the isolated low value of the $N=2$ curve at $U=t/2$ in Fig.~\ref{fig:combined_qsvt}(b). For larger $N$, $\tilde E_0$ starts beyond this zero and the dip is absent. The same mechanism, with the zero of the odd target at $x\approx-0.77$, produces the low $N=2$ value at $U=t$ in Fig.~\ref{fig:combined_qsvt}(a).

\section{Conclusion}

We have investigated the success probability of block-encoding-based quantum algorithms and shown that, for a given encoded function $f$ of a Hermitian operator $H$, it is jointly determined by the spectral distribution of the input state, the spectrum of $H$, and the normalization. 
All numerical tests have been done on the Hubbard model at half filling, in which the spectral structure can be validated and the relevant quantities can be computed over a broad range of system sizes; we expect the mechanisms identified here to hold beyond this case, while the specific values are model dependent.
This provides a common perspective on LCU, qubitization and QSVT block-encodings, in which the success probability can be written as $p^{(f)}_{\mathrm{success}} = \|f(\widetilde{H})\ket{\psi}\|^2 = \sum_n w_n |f(\tilde{E}_n)|^2$, a quantity that can be analyzed independently and included in resource estimation.

For LCU, the success probability is constrained by the squared ratio $(\Omega_s/\lambda_s)^2$ between the spectral norm of the shifted Hamiltonian and the LCU $1$-norm of its Pauli decomposition. Within this bound, the success probability depends strongly on the input state: the half-filled ground state nearly saturates it, whereas the HF input state generally does not.

Qubitization introduces the Chebyshev degree $k$ as an additional control parameter, and the resulting success probability oscillates with $k$. For a fixed block-encoding and input state, different degrees can therefore give substantially different shot overheads. For the HF input state, this degree dependence is largely captured by a small number of dominant spectral components: a spectrum-informed truncation identifies favorable degrees from a few dominant eigenstates, while a spectrum-free model fitted to shallow-degree data predicts favorable deeper degrees without classical diagonalization.

For QSVT, the polynomial scaling factor $C$ provides a different form of control. For a fixed target polynomial and input state, the success probability scales as $1/C^2$, so tighter normalization directly improves it. However, choosing $C$ close to the true supremum $C_\mathrm{opt}$ introduces a practical trade-off: an undershooting value is infeasible, while a barely feasible value places phase finding close to an ill-conditioned boundary. Among the estimates and bounds of $C_\mathrm{opt}$ compared here, the certified bound $C_\mathrm{Markov}$ that we introduce for QSVT normalization, obtained from grid samples through the Markov inequality, provides the best compromise: it stays within about one percent of the tight value while leaving a positive feasibility margin for robust phase synthesis. For the $N=2$ Hubbard chain studied, the QSVT success probability decays with polynomial degree (upper envelope $\propto d^{-1.8}$--$d^{-1.9}$ for the HF input state), nearly independently of the coefficient growth for constant and polynomially growing coefficients, while exponentially growing coefficients prevent it from decaying despite a normalization that grows by orders of magnitude.

Across the three methods, the numerical results show that success probability does not have a single universal dependence on system size, interaction strength, or transformation degree. Instead, for each construction, its behavior reflects the combined effect of spectral structure, normalization and input state. In particular, for the Hubbard systems studied here, no systematic decay of the success probability toward zero is observed with system size (up to 35 qubits for LCU block-encoding), and for the HF input it is quasi-independent of the input state's fidelity to the ground state.

Finally, the application to quantum Krylov illustrates why these observations matter. The moments required by a Krylov calculation are associated with different Chebyshev degrees and therefore with different values of $p^{(T_k)}_{\mathrm{success}}$. Even when the moments are measured without postselection, $p^{(T_k)}_{\mathrm{success}}$ is the squared norm of the corresponding Krylov vector, so degrees with small $p^{(T_k)}_{\mathrm{success}}$ require substantially more shots to be resolved to a given relative precision. Beyond this per-degree shot overhead, the success probability also serves as a construction criterion: selecting the Krylov space from degrees with high success probability improves the conditioning of the overlap matrix and the resulting convergence, at the cost of deeper circuits. The success probability therefore adds a cost that is not captured by gate count, circuit depth or qubit count alone, and should be reported as part of the resource estimation of block-encoding-based algorithms. Amplitude amplification~\cite{brassard_2002} can reduce the number of repetitions from $O(1/p^{(f)}_{\mathrm{success}})$ to $O(1/\sqrt{p^{(f)}_{\mathrm{success}}})$, at the cost of $O(1/\sqrt{p^{(f)}_{\mathrm{success}}})$ additional applications of the block-encoding per run; the success probability therefore remains the quantity that sets the overhead, now in circuit depth as well as in repetitions.

Extending our numerical analysis beyond the one-dimensional Hubbard model, e.g., to molecules and larger many-electron systems, represents a natural way forward. Another important outlook is the application to quantum Krylov subspace methods, and in particular to our recent formulation~\cite{oqkd}, which requires a QSVT polynomial transformation without definite parity. Such a polynomial is typically implemented as a linear combination of its even and odd parts, at the cost of an additional normalization factor that can reduce the success probability by up to a factor of 4; the framework developed here directly quantifies this cost. Finally, the success probability of block-encodings used for solving linear systems~\cite{Harrow_2009,Childs_2017} and partial differential equations~\cite{Childs_2021} remains to be evaluated.

\begin{acknowledgments}
We thank TotalEnergies for permission to publish this work, Keith Gray, Donny Cooper and Harry Zong for providing and maintaining the CPU and GPU computing resources used for the simulations, Henri Calandra and Jean-Patrick Mascomère for insightful discussions and comments, and Hadi Rammal for fruitful discussions and collaboration. This work was supported in part by the Maison du Quantique de Nouvelle-Aquitaine ``HybQuant'', as part of the HQI initiative and France 2030, under the French National Research Agency (ANR) grant ANR-22-PNCQ-0002, by the PEPR Quantique under Grant No. ANR-23-PETQ-0006, and by the EUR Light S\&T (PIA3 Program, Grant No. ANR-17-EURE-0027). Circuit simulations were performed with the Quantum Learning Machine (myQLM) software~\cite{QLM}. The authors used a large language model as an assistant for proofreading and editing parts of the text.
\end{acknowledgments}

\appendix

\section{Readout statistics of the Chebyshev moments}
\label{app:odd_readout}
This appendix derives the statistics of the three readouts of the Chebyshev moments compared in Sec.~\ref{sssec:krylov} and Table~\ref{tab:readouts}: the Hadamard test used in this work and the two readouts of Kirby et al.~\cite{Kirby_2023}. Throughout, an outcome of probability $p$ occurs about $Np$ times in $N$ shots, with a fluctuation of about $\sqrt{Np(1-p)}$, so that estimating $p$ itself to a relative precision $\delta$ requires $N\simeq(1-p)/(p\,\delta^2)$ shots; whether this $1/p$ cost appears depends on what is counted.

\paragraph{Hadamard test.} A test qubit prepared in $\ket{+}$ controls $k$ applications of $W$ on $\ket{\Psi_0}=\ket{\mathrm{PREP}}\otimes\ket{\psi}$, and after a second Hadamard gate, its outcome probabilities satisfy
\begin{equation}
\label{eq:hadamard_test}
    \begin{split}
    p_0-p_1&=\mathrm{Re}\,\bra{\Psi_0}W^k\ket{\Psi_0}\\
    &=\mathrm{Re}\,\bra{\psi}\big(\bra{\mathrm{PREP}}W^k\ket{\mathrm{PREP}}\big)\ket{\psi}
    =m_k ,
    \end{split}
\end{equation}
by Eq.~\eqref{eq:chebyshev}, $m_k$ being real since $T_k(\widetilde H)$ is Hermitian. The projection onto $\ket{\mathrm{PREP}}$ that defines the block-encoding is therefore contained in the overlap with the initial state: the ancilla register is never measured, and every shot, with outcome $\pm1$, contributes to the estimate of $m_k$.

\paragraph{Readouts of Kirby et al.} Both start from the walk state
\begin{equation}
\label{eq:walk_state_kirby}
    \ket{\Psi_j}=W^{j}\big(\ket{\mathrm{PREP}}\otimes\ket{\psi}\big),\qquad j=\lfloor k/2\rfloor,
\end{equation}
and use $\bra{\Psi_j}R_\mathrm{PREP}\ket{\Psi_j}=m_{2j}$ and $\bra{\Psi_j}\mathrm{SELECT}\ket{\Psi_j}=m_{2j+1}$~\cite{Kirby_2023}.
 
\paragraph{Even $k$.} $R_\mathrm{PREP}$ is measured by applying $\mathrm{PREP}^\dagger$ and recording $+1$ if $a=0$ and $-1$ otherwise. The probability of the $+1$ outcome is
\begin{equation}
\label{eq:even_readout}
    \begin{split}
    \Pr(a=0)&=\big\|(\bra{\mathrm{PREP}}\otimes I)\ket{\Psi_{k/2}}\big\|^2\\
    &=p^{(T_{k/2})}_{\mathrm{success}}=\frac{1+m_k}{2},
    \end{split}
\end{equation}
i.e., exactly the block-encoding success probability of the half-degree walk, using $T_{k/2}^2=(1+T_k)/2$. Since $m_k=2\Pr(a=0)-1$, failures are not discarded but carry information on $m_k$, and the error is $\sqrt{(1-m_k^2)/N}$, the same as for the Hadamard test.
 
\paragraph{Odd $k$.} $\mathrm{SELECT}$ is measured term by term~\cite{Kirby_2023}: for term $i$, a layer of single-qubit Clifford gates maps $P_i$ to a diagonal operator, the ancilla and system registers are measured in the computational basis, and the shot returns the measured eigenvalue of $P_i$ if $a=i$, and $0$ otherwise. Only the shots with $a=i$, which occur with probability $q_k(i)$, carry information on term $i$, so the relative error on its contribution scales as $1/\sqrt{N_iq_k(i)}$, where $N_i$ is the number of shots spent on that term. With the optimal allocation $N_i\propto\sqrt{q_k(i)}$, the total cost for a given error on $m_k$ exceeds that of the even readout, in the worst case, by a factor $\big(\sum_i\sqrt{q_k(i)}\big)^2$, which lies between $1$ (a single dominant term) and $L$ (equal weights); grouping qubit-wise commuting Pauli strings reduces $L$ to the number of groups. This overhead is not an artifact of the bounds. Writing $e_i$ for the mean outcome of a shot on term $i$, with $|e_i|\le q_k(i)$ and $m_k=\sum_ie_i$, the variance per shot on term $i$ is $q_k(i)-e_i^2$, whereas that of the Hadamard test and of the even readout is $1-m_k^2$, and
\begin{equation}
\label{eq:odd_vs_hadamard}
    \Big(\sum_i\sqrt{q_k(i)-e_i^2}\Big)^2\ \ge\ 1-m_k^2 ,
\end{equation}
since, after expanding both squares, each cross term satisfies $\sqrt{(q_k(i)-e_i^2)(q_k(l)-e_l^2)}+e_ie_l\ge0$, which follows from $q_k(i)-e_i^2\ge|e_i|(1-|e_i|)$ and $|e_i|+|e_l|\le q_k(i)+q_k(l)\le1$. The odd readout therefore never requires fewer shots than the Hadamard test for the same error on $m_k$, and, since $q_k(i)-e_i^2\ge q_k(i)\big(1-q_k(i)\big)$, it requires at least $L-1$ times more when $q_k(i)=1/L$ for all $i$.
 
The probability $q_k(i)$ follows from the branch structure preserved by the walk, in which each ancilla state $\ket{i}$ carries its own system state. As in SELECT, the signs of the coefficients are absorbed into the Pauli strings, so that $H_s=\sum_i\alpha_iP_i$ with $\alpha_i>0$ and $\sum_i\alpha_i=\lambda_s$. Suppose
\begin{equation}
\label{eq:branch_form}
    \ket{\Psi_j}=\sum_{i}\sqrt{\frac{\alpha_i}{\lambda_s}}\,\ket{i}\otimes\ket{\chi_i^{(j)}},\qquad \ket{\chi_i^{(0)}}=\ket{\psi},
\end{equation}
which holds at $j=0$. $\mathrm{SELECT}$ applies $P_i$ in branch $i$, and projecting onto $\ket{\mathrm{PREP}}$ recombines the branches,
\begin{equation}
    (\bra{\mathrm{PREP}}\otimes I)\,\mathrm{SELECT}\ket{\Psi_j}=\sum_l\frac{\alpha_l}{\lambda_s}\,P_l\ket{\chi_l^{(j)}}.
\end{equation}
Applying $R_\mathrm{PREP}=2\ket{\mathrm{PREP}}\bra{\mathrm{PREP}}-I$ restores the form of Eq.~\eqref{eq:branch_form}, with
\begin{equation}
\label{eq:branch_recursion}
    \ket{\chi_i^{(j+1)}}=2\sum_l\frac{\alpha_l}{\lambda_s}\,P_l\ket{\chi_l^{(j)}}-P_i\ket{\chi_i^{(j)}}.
\end{equation}
By the Born rule, the probability of reading the ancilla in $\ket{i}$ for odd $k$ is
\begin{equation}
\label{eq:odd_Pk}
    q_k(i)=\frac{\alpha_i}{\lambda_s}\,\big\|\chi_i^{(j)}\big\|^2,\qquad j=\frac{k-1}{2}.
\end{equation}
Using $\sum_l(\alpha_l/\lambda_s)P_l=\widetilde H$, $\sum_l\alpha_l/\lambda_s=1$ and $P_l^2=I$, Eq.~\eqref{eq:branch_recursion} is solved by 
\begin{equation}
\label{eq:odd_closed_form}
    \ket{\chi_i^{(j)}}=\Big[U_j(\widetilde H)-P_i\,U_{j-1}(\widetilde H)\Big]\ket{\psi},
\end{equation}
where $U_{-1}=0$, $U_0=1$ and $U_{j+1}(x)=2x\,U_j(x)-U_{j-1}(x)$ are the Chebyshev polynomials of the second kind, as in Eq.~\eqref{eq:pprime}. For $k=1$, $q_1(i)=\alpha_i/\lambda_s$ exactly. For $k=3$,
\begin{equation}
\label{eq:odd_k3}
    \begin{split}
    q_3(i)&=\frac{\alpha_i}{\lambda_s}\,\big\|(2\widetilde H-P_i)\ket{\psi}\big\|^2\\
    &=\frac{\alpha_i}{\lambda_s}\Big[1+4\langle\widetilde H^2\rangle-2\langle\widetilde HP_i+P_i\widetilde H\rangle\Big],
    \end{split}
\end{equation}
and, using $\sum_i(\alpha_i/\lambda_s)P_i=\widetilde H$, $\sum_iq_3(i)=1+4\langle\widetilde H^2\rangle-4\langle\widetilde H^2\rangle=1$. For the illustrative case $\widetilde H=0.8\,Z+0.2\,X$ and $\ket{\psi}=\ket{0}$, the probability of the $Z$ branch decreases from $0.80$ to $0.42$ and $0.14$ for $k=1,3,5$, while that of the $X$ branch increases accordingly.

\bibliography{prxbib}

\end{document}